\documentclass[aps,prd,nofootinbib,twocolumn,superscriptaddress,preprintnumbers,balancelastpage,longbibliography]{revtex4-1}

\usepackage{appendix}
\usepackage[utf8]{inputenc}
\usepackage{amsmath,amssymb,mathtools,bm}
\usepackage{graphicx, color, hepunits}
\usepackage[dvipsnames]{xcolor}
\usepackage{float}
\usepackage{multirow}
\usepackage{tikz-feynman} 
 \usepackage{hyperref}
\hypersetup{
    colorlinks=true,       
    linkcolor=blue,        
    citecolor=blue,        
    filecolor=magenta,     
    urlcolor=blue          
}
\usepackage[utf8]{inputenc}
\usepackage[english]{babel}

\usepackage{tensor}

\newcommand{\g}{g_{aNN}}
\newcommand{\He}{^{3}{\rm He}}

\begin{document}

\title{The statistics and sensitivity of axion wind detection with the homogeneous precession domain of superfluid helium-3}

\author{Joshua W. Foster}
\affiliation{Center for Theoretical Physics, Massachusetts Institute of Technology, Cambridge, Massachusetts 02139, U.S.A}

\author{Christina Gao}
\affiliation{Department of Physics, University of Illinois Urbana-Champaign, Urbana, IL 61801, USA}
\affiliation{Illinois Center for Advanced Studies of the Universe,
University of Illinois Urbana-Champaign, Urbana, IL 61801, USA}
\affiliation{Theoretical Physics Division, Fermi National Accelerator Laboratory, Batavia, IL 60510, USA}
\affiliation{Superconducting Quantum Materials and Systems Center (SQMS), Fermi National Accelerator Laboratory, Batavia, IL 60510, USA}

\author{William Halperin}
\affiliation{Department of Physics and Astronomy, Northwestern University, Evanston, IL 60208,USA}
\affiliation{Superconducting Quantum Materials and Systems Center (SQMS), Fermi National Accelerator Laboratory, Batavia, IL 60510, USA}

\author{Yonatan Kahn}
\affiliation{Department of Physics, University of Illinois Urbana-Champaign, Urbana, IL 61801, USA}
\affiliation{Illinois Center for Advanced Studies of the Universe,
University of Illinois Urbana-Champaign, Urbana, IL 61801, USA}
\affiliation{Superconducting Quantum Materials and Systems Center (SQMS), Fermi National Accelerator Laboratory, Batavia, IL 60510, USA}

\author{Aarav Mande}
\affiliation{Department of Physics, University of Illinois Urbana-Champaign, Urbana, IL 61801, USA}

\author{Man Nguyen}
\affiliation{Department of Physics and Astronomy, Northwestern University, Evanston, IL 60208,USA}

\author{Jan Sch\"utte-Engel}
\affiliation{Department of Physics, University of Illinois Urbana-Champaign, Urbana, IL 61801, USA}
\affiliation{Illinois Center for Advanced Studies of the Universe,
University of Illinois Urbana-Champaign, Urbana, IL 61801, USA}
\affiliation{University of California, Berkeley, CA 94720, USA}
\affiliation{RIKEN iTHEMS, Wako, Saitama 351-0198, Japan}

\author{John William Scott}
\affiliation{Department of Physics and Astronomy, Northwestern University, Evanston, IL 60208,USA}

\date{\today}
\preprint{MIT-CTP/5601}
\preprint{RIKEN-iTHEMS-Report-23}

\begin{abstract}
The homogeneous precession domain (HPD) of superfluid $^{3}$He has recently been identified as a detection medium which might provide sensitivity to the axion-nucleon coupling $g_{aNN}$ competitive with, or surpassing, existing experimental proposals. In this work, we make a detailed study of the statistical and dynamical properties of the HPD system in order to make realistic projections for a full-fledged experimental program. We include the effects of clock error and measurement error in a concrete readout scheme using superconducting qubits and quantum metrology. This work also provides a more general framework to describe the statistics associated with the axion gradient coupling through the treatment of a transient resonance with a non-stationary background in a time-series analysis. Incorporating an optimal data-taking and analysis strategy, we project a sensitivity approaching $g_{aNN} \sim 10^{-12}$ GeV$^{-1}$ across a decade in axion mass. 
\end{abstract}
\maketitle

\section{Introduction}

Axions and axion-like particles are well-motivated candidates for dark matter (DM)~~\cite{Preskill:1982cy,Abbott:1982af,Dine:1982ah}, and their phenomenology is characterized by sub-eV masses and weak couplings to photons and/or Standard Model fermions~\cite{Sikivie:1983ip}. The shift symmetry of the axion field $a$ (arising from its origin as a pseudo-Goldstone boson) implies that its couplings to fermions must involve a derivative; in the non-relativistic regime relevant for DM, this becomes a gradient coupling to nuclear spins $\bm{\sigma}_n$ in the Hamiltonian~\cite{Graham:2013gfa},
\begin{equation}
H \supset \g \nabla a \cdot \bm{\sigma}_n,
\end{equation}
where $\g$ is a coupling constant with dimensions of inverse energy in natural units. 

Since the form of the Hamiltonian is identical to the way an ordinary magnetic field couples to spins, $H \supset \gamma \mathbf{B} \cdot \bm{\sigma}_n$, it is often convenient to think of this ``axion wind'' coupling as an effective magnetic field
\begin{equation}
\label{eq:Ba}
\mathbf{B}_a \equiv \frac{\g}{\gamma} \nabla a,
\end{equation}
where $\gamma$ is the gyromagnetic ratio of the nucleus in question. There have been several theoretical ideas and a number of recent experimental results aiming to detect this coupling by exploiting nuclear magnetic resonance~\cite{Graham:2013gfa,Garcon:2017ixh,Graham:2017ivz,JacksonKimball:2017elr,Wu:2019exd,Garcon_2019,Aybas:2021nvn,Jiang:2021dby,Aybas:2021cdk,Bloch:2021vnn,Aybas:2021nvn,Lee:2022vvb,Bloch:2022kjm,Dror:2022xpi}, including a recent proposal to use the homogeneous precession domain (HPD) of superfluid $\He$ in the B-phase~\cite{Gao:2022nuq}.\footnote{See Refs.~\cite{Irastorza:2018dyq,Adams:2022pbo} for a review of other experimental approaches, and Refs.~\cite{Kakhidze:1991,Arvanitaki:2017nhi,Lawson:2019brd,Chigusa:2020gfs,Mitridate:2020kly,Schutte-Engel:2021bqm,Chigusa:2021mci,Roising:2021lpv,Arvanitaki:2021wjk,Berlin:2022mia,Hochberg:2022apz,Hong:2022nss,Marsh:2022fmo,Shaposhnikov:2023pdj,Berlin:2023ppd,Chigusa:2023hmz} for proposals which exploit particular properties of condensed matter systems.} The HPD can be understood as a Bose condensate of magnons~\cite{bunkov2012spin}, and its key feature is a linearly drifting Larmor precession frequency $\omega_L$ of the sample, which allows a broadband scan over candidate axion masses $m_a$.

In this paper, we extend the results of~\cite{Gao:2022nuq} by including a full statistical treatment of the axion field. Our statistical treatment is similar to that of \cite{Lisanti:2021vij, Gramolin:2021mqv}, but differs in that we develop complete time domain statistics of the axion gradient field as a stochastic force driving a nonstationary dynamical system. Due to its low mass and hence large occupation number, $a$ and its gradient may be treated as a Gaussian random field characterized by its two-point correlation function. Using this fact, we construct a likelihood function for the hypothetical axion signal in terms of time-domain correlation functions of the precessing HPD magnetization signal evolving under the Bloch equations, and we incorporate the statistics of the stochastically-fluctuating background as well. We find that additional signal-to-noise information may in principle be extracted by considering the HPD signal well after the resonance condition $\omega_L = m_a$, due to the distinctive beat frequencies between the drifting precession frequency and the effective axion magnetic field, though this requires fast measurements of the instantaneous precession frequency which may be difficult to realize in practice. A likelihood analysis leads to practical prescriptions for optimizing experimental parameters such as the magnetic field gradient and data-taking procedures such as the frequency sampling rate, which go beyond the initial analysis of~\cite{Gao:2022nuq}. In performing such an optimization, we specify a concrete scheme for measuring the HPD precession frequency (or equivalently, the precession drift $\dot{\omega}$) using the techniques of quantum metrology and optimal quantum control~\cite{pang2017optimal,naghiloo2017achieving}, where we imagine that the signal is read out using an array of superconducting transmon qubits time-stamped with a high-precision atomic clock~\cite{boss2017quantum,schmitt2017submillihertz}.

This paper is organized as follows. In Sec.~\ref{sec:GradientStats}, we review the statistics of the axion gradient field as a necessary input for our calculations, with further details included in Appendix~\ref{app:Gaussian}. In Sec.~\ref{sec:Dynamics}, we evaluate in detail the dynamics and statistics of the HPD system evolving under the Bloch equations in the presence of a spatially-varying magnetic field and stochastic axion gradient. In Sec.~\ref{sec:StochasticDynamics}, we characterize the statistics of the background evolution of the precession frequency due to irreducible noise from stochastic magnon loss; we join these statistics with those of the axion field to develop a full statistical characterization of HPD system and likelihood framework for projected sensitivities and future analyses. In Sec.~\ref{sec:Sensitivity}, we evaluate the HPD sensitivity to the axion coupling accounting for realistic clock noise and measurement noise in addition to the system's stochastic dynamics. We determine projected sensitivities for various experimental configurations, including a strategy for collecting data suitable for the analysis we construct. In Sec.~\ref{sec:Modulation}, we also consider the impact of daily and annual modulation for sensitivities. Finally, we provide some concluding remarks in Sec.~\ref{sec:Conclusion}.

A particularly interesting feature of HPD experiments is that, though they can be characterized via an effective macroscopic description through the Bloch equations, details such as the smooth precession frequency drift appear deterministic only in the limit of coarse-graining the stochastic loss of magnons in the condensate. These stochastic fluctuations are the irreducible limiting background for axion searches. As a result, the dynamics of the HPD -- with and without an axion gradient field source -- are intrinsically stochastic. In Appendix~\ref{App:SDE}, we study this feature of the system within the framework of stochastic differential equations, finding the impact of the microscopic stochastic dynamics to be negligible.

\section{Statistics of the Axion Gradient Field}
\label{sec:GradientStats}

In this section, we determine the time-domain statistics of axion gradient field relevant for calculating the effect of axion DM on the evolution of the HPD. The statistics of the gradient field have been previously studied in several contexts, including \cite{Lisanti:2021vij, Lee:2022vvb, Dror:2022xpi, Gramolin:2021mqv}, but our treatment here differs in that we develop general two-point correlators in time between components of the gradient field rather than making an equivalent treatment in the frequency domain or considering time-time correlators in more directly observable quantities. In particular, the two-point correlators in time which we develop are the ones most relevant for calculating the statistics of transient axion-induced shift in the precession frequency, whereas previous works considered stationary processes. 

We begin with a simple construction of a discretized realization of the axion field, as in \cite{Foster:2020fln}:
\begin{equation}
\begin{split}
    a(\mathbf x, t) = \frac{\sqrt{\rho_a}}{m_a} \sum_{abc}& \alpha_{abc} \sqrt{f(\mathbf{v}_{abc}) (\Delta^3 v_{abc})} \\
    & \times \cos\left[\omega_{abc} t - \mathbf{k}_{abc}\cdot \mathbf{x} + \phi_{abc}\right]
\end{split}
\end{equation}
where $\rho_a \simeq 0.4 \ {\rm GeV}/{\rm cm^3}$ is the DM energy density, $abc$ is a multi-index for a 3-dimensional discretization of the DM velocity $\mathbf{v}_{abc}$ in small volumes of size $\Delta^3 v_{abc}$, $f(\mathbf{v})$ is the DM velocity distribution, $\omega_{abc} = m_a (1 + \mathbf{v}_{abc}^2/2)$ is the DM frequency, and $\mathbf{k}_{abc} = m_a \mathbf{v}_{abc}$ is the wavenumber. The stochastic nature of $a$ is controlled by the random variables $\alpha_{abc}$ and $\phi_{abc}$, which are Rayleigh-distributed on $[0, \infty)$ and uniformly-distributed on $[0, 2\pi)$, respectively.\footnote{Note that this construction is identical to that in \cite{Foster:2017hbq}, except now the field is being constructed by integration over the full three-dimensional velocity distribution, rather than the one-dimensional speed distribution.} The axion gradient evaluated at a single spatial point follows immediately as 
\begin{equation}
\begin{split}
    \nabla a(t) =  \sqrt{\rho_a} \sum_{abc}& \alpha_{abc} \sqrt{f(\mathbf{v}_{abc}) (\Delta^3 v_{abc})} \\
    & \times \cos\left[\omega_{abc} t + \phi_{abc}\right] \mathbf{v}_{abc}
\end{split}
\end{equation}
where we have absorbed any dependence on the chosen position $\mathbf{x}$ into the uniformly distributed phase.
As it is constructed from summing uncorrelated plane waves with uniformly-distributed phases, $\nabla a(t)$ is a Gaussian process, and hence its statistics are fully characterized by its one- and two-point correlation functions. We proceed to evaluate those defining moments. In Appendix.~\ref{app:Gaussian}, we present Monte Carlo tests which validate their expected Gaussianity using a treatment adapted from \cite{Foster:2017hbq} and utilized in similar context for frequency domain statistics of axion gradient signals in \cite{Gramolin:2021mqv}.

Evaluating the first moment of $\nabla a$ is actually trivial. Since each phase is uncorrelated and drawn uniformly on $[0,\, 2\pi)$, the expectation value of the contribution of any plane wave mode to $\nabla a$ is zero, and thus
\begin{equation}
    \langle \nabla_i a  \rangle = 0.
\end{equation}
Determining the covariance is somewhat more involved. Working with our discrete realization, 
\begin{widetext}
\begin{equation}
\begin{split}
    \Sigma_{ij}(t, t') \equiv \langle \nabla_i a(t) \nabla_j a(t') \rangle = \langle \rho_a \sum_{abcpqr} &\alpha_{abc} \alpha_{pqr} \sqrt{f(\mathbf{v}_{abc})f(\mathbf{v}_{pqr}) (\Delta^3 v_{abc})  (\Delta v_{pqr})^3 } \\
    &\times \cos\left[\omega_{abc} t + \phi_{abc} \right] \cos\left[\omega_{pqr} t' + \phi_{pqr}\right] \mathbf{v}^i_{abc} \mathbf{v}^j_{pqr} \rangle.
\end{split}
\end{equation}
\end{widetext}
Since $\phi_{abc}$ and $\phi_{pqr}$ are uniform on $[0, 2\pi)$ and uncorrelated, the summand vanishes in the expectation value unless the multi-indices are equal, $abc = pqr$. After simplifying trigonometric terms and keeping only those with nonzero expectation value, we obtain
\begin{equation}
\begin{split}
    \Sigma_{ij}(t, t') = \frac{\rho_a}{2}  \sum_{abc}& (\Delta^3 v_{abc}) \mathbf{v}^i_{abc} \mathbf{v}^j_{abc} f(\mathbf{v}_{abc}) \\
    &\times \cos\left[\omega_{abc} (t - t') \right] \langle  \alpha_{abc}^2  \rangle.
\end{split}
\end{equation}
For a Rayleigh-distributed variable $\alpha_{abc}$, $\langle \alpha_{abc}^2 \rangle = 2$, so the covariance reduces to
\begin{equation}
    \Sigma_{ij}(t, t') =  \rho_a  \sum_{abc} (\Delta^3 v_{abc}) \mathbf{v}^i_{abc} \mathbf{v}^j_{abc} f(\mathbf{v}_{abc}) \cos\left[\omega_{abc} (t - t') \right].
\end{equation}
We can now take the continuum limit $\Delta^3 v_{abc} \to d^3 \, \mathbf{v}$ and $\omega_{abc} \to \omega_{\mathbf{v}} \equiv m_a (1 + \mathbf{v}^2/2)$ to obtain our final form for the covariance,
\begin{gather}
\label{eq:GradCov}
    \Sigma_{ij}(t, t') = \rho_a \int dv \, F_{ij}(v) \cos\left[\omega_\mathbf{v} (t - t') \right].
\end{gather}
where we have defined for future notational convenience
\begin{equation}
    F_{ij}(v) \equiv \int d\Omega \, \mathbf{v}^2 \mathbf{v}_i \mathbf{v}_j f(\mathbf{v}),
\end{equation}
with $d\Omega \, \mathbf{v}^2$ the measure on the unit sphere in velocity space. We note in passing that the axion field is often characterized by a coherence time defined along the lines of 
\begin{equation}
    \tau \sim \frac{2\pi}{m_a v^2} \sim 80 \ {\rm ms} \times \left(\frac{50 \ {\rm neV}}{m_a}\right)
    \label{eq:CoherenceTime}
\end{equation}
where $v \sim 300 \ {\rm km/s}$ is a typical DM speed. Use of a coherence time in our subsequent calculations is unnecessary, as it serves as merely a proxy for the full covariance structure encapsulated in Eq.~(\ref{eq:GradCov}).

\section{Dynamics of the HPD with an axion source}
\label{sec:Dynamics}
\begin{figure*}[htb]  
\hspace{0pt}
\vspace{-0.2in}
\begin{center}
\includegraphics[width=0.99\textwidth]{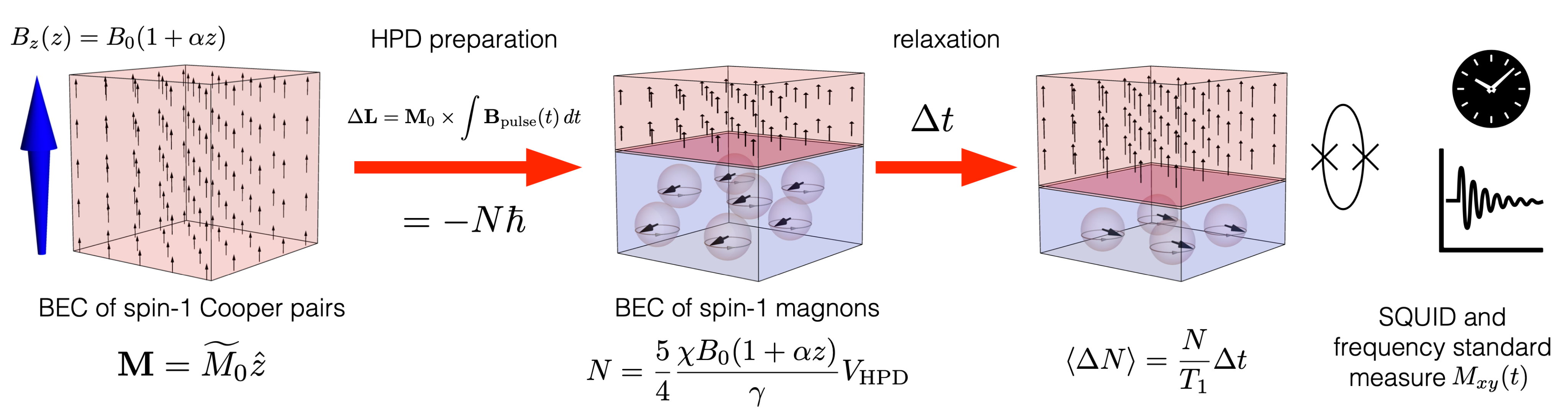}
\caption{Preparation and evolution of the HPD. Superfluid $\He$-B is subjected to an external magnetic field with a gradient along the field direction, establishing an equilibrium magnetization (\emph{left}), Eq.~(\ref{eq:M0z}). The HPD is prepared by injecting angular momentum into the system via a transverse magnetic field pulse; the resulting spin-1 magnons Bose-condense into the HPD (\emph{center}), with volume $V_{\rm HPD}$ determined by the total number of magnons $N$ as in Eq.~(\ref{eq:TotalMagnonNumber}). The spins in the HPD (blue) precess with a tip angle $\beta_0 =104^\circ$ at a frequency determined by the magnetic field $B(z)$ at the domain wall between the HPD and the relaxed phase (Eq.~(\ref{eq:Larmorbkg})), while the remainder of the sample (red) remains at its equilibrium magnetization. As time elapses (\emph{center right}), magnon losses (Eq.~(\ref{eq:avgmagnonloss})) lead to a relaxation of the HPD on a time scale $T_1$, shrinking the HPD volume and causing the precession frequency to drift as the domain wall moves downward. The time-dependent transverse magnetization is sensed by a SQUID referenced to a frequency standard (\emph{right}) to enable the optimal quantum control measurement of the precession frequency drift; see Sec.~\ref{sec:Sensitivity} for more details.}
\label{fig:HPDCombined}
\end{center}
\end{figure*}

As a Bose-Einstein condensate (BEC) of spin-1 magnons, the dynamics of the HPD are intrinsically stochastic and quantum mechanical in nature. However, as a macroscopic quantum system, the HPD dynamics can largely be characterized by considering its effective description via the Bloch equations, as was done in \cite{Gao:2022nuq}. In this section, we carefully account for the effects of the axion gradient coupling in the Bloch description, before returning in the following Sec.~\ref{sec:StochasticDynamics} to the magnon picture that allows for a much more detailed accounting of background processes and their contribution to statistical uncertainties. Fig.~\ref{fig:HPDCombined} illustrates the preparation of the HPD, background evolution of the HPD in the absence of an axion, and measurement of the HPD.

\subsection{Effective Description in the Bloch Equations}

We begin with the Bloch equations developed in \cite{Gao:2022nuq} that describe the evolution of the HPD magnetization $\mathbf{M}$ subject to an external (possibly time-varying and inhomogeneous) magnetic field $\mathbf{B}$. Here, the magnetization $\mathbf{M}$ is defined as $\mathbf{M} = \mathbf{m}/V_\mathrm{tot}$ where $\mathbf{m}$ is the HPD magnetic moment and $V_\mathrm{tot}$ is the total $\He$ volume (including both the HPD and the relaxed domain). Assuming an equilibrium magnetization $\widetilde{M_0}$ in the $z$-direction, the parallel magnetization $M_z$ and transverse magnetizations $M_x$ and $M_y$ evolve as
\begin{gather}
\label{eq:Bloch}
    \dot M_z = \frac{i \gamma}{2} \left(M_{xy} \bar B_{xy} - \bar{M}_{xy}B_{xy} \right) - \frac{M_z- \widetilde{M_0}}{5 T_1} \\ 
    \dot M_{xy} = - i \gamma (M_{xy} B_z - M_z B_{xy}) - \frac{M_{xy}}{T_1}
\end{gather}
where $M_{xy} = M_x + i M_y$,  $B_{xy} = B_x + i B_y$, and $T_1$ is the characteristic relaxation time of the system. We will consider a system of total height $h$ with cross-sectional area $A$ with the magnetic field pointing in the $\hat z$ direction varying as 
\begin{equation}
B_z(z) = B_0 (1 + \alpha z),
\end{equation}
and convention, we will take $\alpha > 0$ and $B_0 > 0$ to be the background magnetic field strength at $z = 0$. 

The equilibrium magnetization of the HPD system is given by 
\begin{equation}
    \widetilde{M_0} = \chi B \mathbb{F}
\end{equation}
where $\chi\sim 10^{-7}$ is the magnetic susceptibility for $^3$He~\cite{Wheatley_1975}, $\mathbb{F}$ is the fraction of the sample in the HPD phase, and $B$ is the field strength in the $\hat z$ direction at the location of the domain wall which separates the relaxed and precessing phases. For our geometry, the HPD fraction is $\mathbb{F} = z/h$ where $z$ is the position of the domain wall. Assuming the initial transverse pulse is sufficient for the HPD to initially encompass the whole sample, the domain wall will descend from the top of the container $z = h$ at $t = 0$ to $z = 0$ as $t\rightarrow\infty$. Then we can write 
\begin{equation}
    \widetilde M_0(z) = \chi B_0 (1 + \alpha z) \frac{z}{h},
    \label{eq:M0z}
\end{equation}
suggesting that the domain wall height $z$ is a natural variable to use in the Bloch equations. 

We choose the transverse magnetization phase $\theta$ as the other variable. As noted in Ref.~\cite{Gao:2022nuq}, since the HPD features a transverse and longitudinal magnetization that are locked together at the Leggett angle $\beta_0 = \cos^{-1}(-1/4) \approx 104^\circ$, $M_z$ and $M_{xy}$ do not evolve independently, and thus $z$ and $\theta$ are sufficient to fully characterize the evolution of the HPD.

In the presence of axion DM, the HPD couples to the effective axion magnetic field $\mathbf{B}_a(t)$ in Eq.~(\ref{eq:Ba}), which will affect the dynamics of the HPD system in Eq.~(\ref{eq:Bloch}) but leaves the equilibrium magnetization $\widetilde{M_0}$ unchanged. Since the axion coupling is small, it is then natural to expand our dynamical variables $z(t)$ and $\theta(t)$ perturbatively to first order in $g_{aNN}$ so that $z(t) \approx z_0(t)(1 + z_a(t))$ and $\theta(t) \approx \theta_0(t) + \theta_a(t)$. Note that we have defined $z_a$ as a dimensionless shift in the domain wall location relative to the free evolution $z_0$.

In terms of our dynamical variables, we may write the components of the magnetization as
\begin{equation}
\begin{split}
    M_x(t) &= \widetilde M_0\bigg(z_0(t)(1+z_a(t))\bigg) \sin \beta_0 \cos[\theta_0(t) + \theta_a(t)], \\
    M_y(t) &= \widetilde M_0\bigg(z_0(t)(1+z_a(t))\bigg) \sin \beta_0 \sin[\theta_0(t) + \theta_a(t)], \\
    M_z(t) &= \widetilde M_0\bigg(z_0(t)(1+z_a(t))\bigg)  \cos \beta_0.
\end{split}
\end{equation}
These magnetizations evolve subject to a magnetic field
\begin{align}
    \mathbf{B} &= \begin{bmatrix}
            B_x^a(t) \\
           B_y^a(t) \\
           B_0 [1 + \alpha z_0(t)(1 + z_a(t))] + B_z^a(t)
         \end{bmatrix}.
\end{align}
where $B^a_{i}$ are the components of the axion effective magnetic field $\mathbf{B}_a$ defined in Eq.~(\ref{eq:Ba}). 
Substituting these quantities into the Bloch equations and working to zeroth-order in axion coupling, we obtain the equations of motion for the background evolution of the HPD system:
\begin{gather}
\dot z_0(t) = -\frac{z_0(t) \left(\alpha  z_0(t)+1\right)}{T_1 \left(2 \alpha  z_0(t)+1\right)}, \\ 
\theta_0(t) = \gamma B_0 \left[1 +\alpha  z_0(t)\right],
\label{eq:theta0}
\end{gather} For completeness, we provide the solution
\begin{equation}
    z_0(t) = \frac{e^{-\frac{t}{2 T_1}} \sqrt{e^{\frac{t}{T_1}}+4 \alpha  h (\alpha h+1)}-1}{2 \alpha }
\end{equation}
for the domain wall height with boundary condition $z_0(t= 0) = h$. An analytic solution for $\theta_0$ is straightforwardly obtained by substituting into Eq.~(\ref{eq:theta0}).

Next, the equation of motion at first-order in the axion coupling for the axion-induced domain wall motion is
\begin{equation}
\begin{split}
\dot z_a(t) = & \bigg[B^a_x(t) \sin \left( \theta_0(t)\right)+B^a_y(t) \cos \left(\theta _0(t)\right)\bigg] \\
& \times \gamma \tan\beta_0 \frac{1+\alpha  z_0(t)}{1+2 \alpha  z_0(t)},
\label{eq:zEoM}
\end{split}
\end{equation}
and the shift it induces in the precession rate is given by 
\begin{equation}
    \dot \theta_a(t) = \alpha \gamma B_0  z_0(t) z_a(t).
\end{equation}
In deriving these equations, we have neglected slowly varying terms which are suppressed by $1/T_1$ since already we are already at first-order in the small coupling parameter. We have also neglected bare oscillatory terms of the form $\sin [\theta(t)]$ or $\cos[\theta(t)]$ that average to zero over intervals longer than the precession period. These terms include those proportional to $B^a_z$, so the longitudinal component of the axion gradient does not contribute to the HPD dynamics over timescales longer than the precession period. 

Thus, by expanding perturbatively, we have then decoupled our two variables, such that after solving for $z_0$ and $z_a$, we may always compute $\omega_L(t) \equiv \dot \theta(t) = \dot{\theta}_0(t) + \dot{\theta}_a(t)$, which is the experimentally-measurable observable of interest. We note that expressing the precession rate as the sum of the precession rates associated with the background and axion-induced domain wall motion is an artifact of our order-by-order expansion. The precession can more simply be written as
\begin{equation}
    \omega_L(t) = \gamma (1 + \alpha z) B_0,
    \label{eq:Larmorbkg}
\end{equation}
where $z = z_0(1+z_a)$ is the domain wall position.

\subsection{Axion Resonance in the Bloch Equations}
\label{sec:ResStats}
We now consider the dynamics of the resonance in our Bloch treatment. We define the time of the resonance $t_r$ by $\omega_L(t_r) = m_a$, where $\omega_L \equiv \dot{\theta}(t)$. We additionally denote the precession frequency derivative at time $t_r$ by $\dot \omega_r$. We may then expand the background phase $\theta_0$ around the time of the resonance by 
\begin{equation}
    \theta_0(t) \approx \theta_0(t_r) + m_a(t-t_r) - \frac{\dot \omega_r}{2}(t - t_r)^2. 
\end{equation}
For definiteness and convenience, we take $\theta_0(t_r) = 0$, which sets the orientation of the magnetization with respect to the Earth's peculiar velocity at the time of the resonance, though this need not be true in a particular experimental realization. We can substitute this expansion into our equation for $\dot z_a(t)$, but first it is informative to consider the statistics of $z_a$. 

\begin{figure*}[htb]  
\hspace{0pt}
\vspace{-0.2in}
\begin{center}
\includegraphics[width=0.99\textwidth]{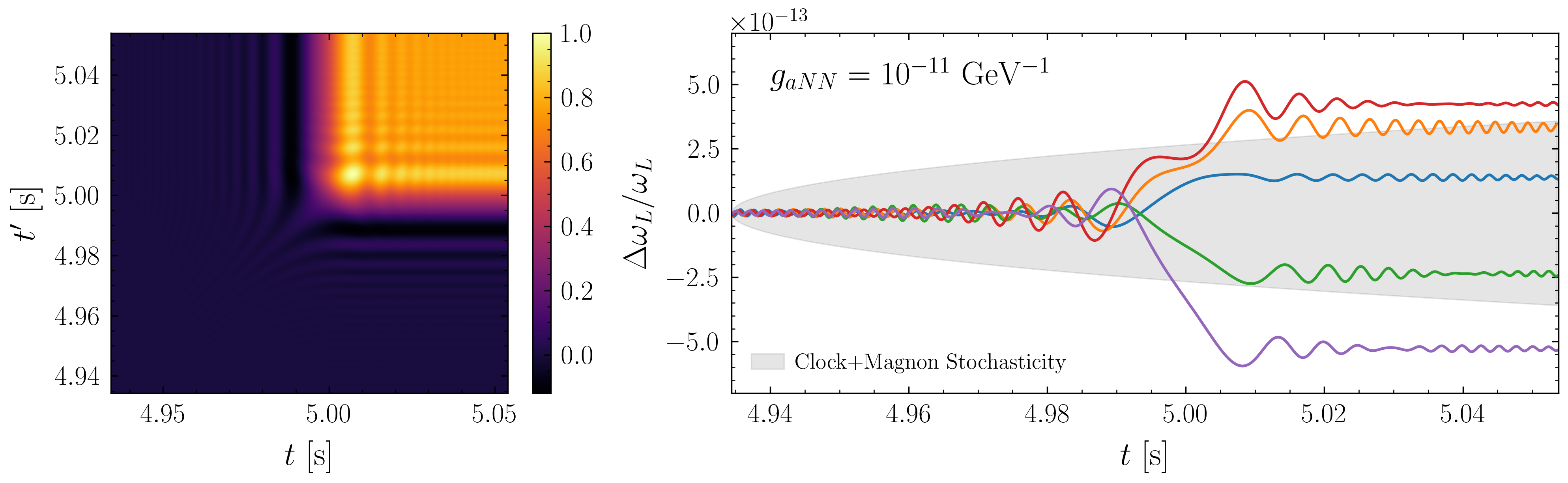}
\caption{Behavior of the axion-induced signal in the HPD for times near the resonance, taken to be $t_r \approx 5\,\mathrm{s}$, for representative parameters $m_a \approx \, 70 \, \mathrm{neV}$, $\alpha = 1\,\mathrm{cm}^{-1}$, $h = 10\,\mathrm{cm}$, $T_1 = 1000 \, \mathrm{s}$, and $B_0 = 0.05\,\mathrm{T}$. (\textit{Left}) The time-dependent covariance of the axion-induced domain wall shift, Eq.~(\ref{eq:zzcorr}), normalized with respect to the maximum variance in the relative displacement of the domain wall. The behavior is approximately that of a step function modulated by beat frequency oscillations. (\textit{Right}) Five realizations of the axion-induced domain wall shift drawn from the multivariate Gaussian distribution specified by the covariance matrix for $g_{aNN} = 10^{-11}$ GeV$^{-1}$. The horizontal grey band illustrates the variance of the measured $\Delta \omega_L / \omega_L$ associated with stochastic magnon loss and clock error. For details, see Sec.~\ref{sec:Sensitivity}.}
\label{fig:SigCovExample}
\end{center}
\end{figure*}

From Eq.~(\ref{eq:zEoM}), $\dot z_a(t)$ depends linearly on $\mathbf{B}_a \equiv (g_{aNN}/\gamma) \nabla a$. As we have established, $\nabla a$ follows a multivariate Gaussian distribution, so $\dot z_a(t)$, which is constructed from a weighted sum over components of $\nabla a$, must be a Gaussian variate itself. Moreover, $z_a$, which is the time integral of Gaussian variates, must also be a Gaussian variate.\footnote{Recall that in this section, we are treating the background evolution $z_0$ and $\theta_0$ as deterministic; see Sec.~\ref{sec:StochasticDynamics} and App.~\ref{App:SDE} for the effects of stochasticity.} Direct integration of $\dot z_a$ then allows us to evaluate the defining expectation values for $z_a(t)$.

Since $\langle \nabla a \rangle = 0$, the mean of $z_a$ vanishes as well:
\begin{equation}
    \langle z_a(t) \rangle = 0.
\end{equation}
We can also calculate the 2-point correlator
\begin{widetext}
\begin{equation}
\label{eq:zzcorr}
\begin{split}
\langle z_a(t) z_a(t') \rangle =  &\tan^2 \beta_0 \, g_{aNN}^2 \, \rho_a \int dv \int_0^t d\tilde t \int_0^{t'} d\tilde t' \cos(\omega_\mathbf{v} (\tilde t - \tilde t')) \frac{1+\alpha  z_0(\tilde t)}{1+2 \alpha  z_0(\tilde t)} \frac{1+\alpha  z_0(\tilde t')}{1+2 \alpha  z_0(\tilde t')} \times \\
&\bigg[F_{xx}(v)\sin \theta_0(\tilde{t}) \sin\theta_0(\tilde t')+ 2 F_{xy}(v) \cos\theta_0(\tilde{t}) \sin\theta_0(\tilde t')+F_{yy}(v) \cos\theta_0(\tilde{t})\cos\theta_0(\tilde{t}')\bigg],
\end{split}
\end{equation}
\end{widetext}
where we have used Eq.~(\ref{eq:GradCov}) to compute the 2-point correlator of $B^a$. While Eq.~(\ref{eq:zzcorr}) is not generally analytically tractable, it can be readily computed numerically. Since $\langle z_a \rangle = 0$, Eq.~(\ref{eq:zzcorr}) also gives covariance of the axion-induced relative shift in the domain wall position.

In Fig.~\ref{fig:SigCovExample}, we plot the covariance in Eq.~(\ref{eq:zzcorr}) for a representative set of values of the experimental parameters, as well as several time-series realizations of such a signal. Two notable features that are highly relevant to the analysis scheme we will subsequently develop are readily apparent. First, since the signal has zero mean, $\langle z_a(t) \rangle  = 0$, an axion search is necessarily a search for extra covariance above the expected background scatter. Second, while the resonant period does indeed induce large motion in the domain wall position, with a covariance that grows quadratically with time, the axion can still drive oscillations in the domain wall position off-resonance. These oscillations can have amplitude on the same order of magnitude as the total resonant shift. Incorporating these oscillatory features in the covariance and expected signal is critical for maximizing the sensitivity of an analysis.

\section{Stochastic Domain Wall Motion}
\label{sec:StochasticDynamics}

While the Bloch description captures the expected coarse-grained evolution of the HPD system, it does not account for the stochasticity of the dissipation processes that lead to magnon loss~\cite{fomin_separation_1996}. This is of critical importance as it was estimated in \cite{Gao:2022nuq} that stochastic magnon loss represented the dominant noise floor for the search. We now adapt our calculations above to account for the microphysical description of the HPD in terms of magnon statistics in order to precisely quantify its effect on our sensitivity.

\subsection{Domain Wall Motion from Stochastic Magnon Loss}

The HPD of $^3$He can be described as a BEC of $N_i$ magnons, with $N_i$ jointly determined by $V_\mathrm{HPD}$ and the B-field at the domain wall location. Given $N_i$ magnons in the system at time $t_{i}$, the expected number of magnons lost over a small time interval $\Delta t$ will be $N_i \Gamma \Delta t$ where $\Gamma =  T_1^{-1}$. Hence, whatever the source of magnon loss, the evolution of magnon number is a Markov process with Poisson statistics in each step. Because the number of magnons in the system is macroscopically large, $N_i \sim 10^{20}$ for the experimental parameters we will consider, the expected number of magnons lost even for $\Delta t$ on the order of milliseconds is sufficiently large to allow for a Gaussian approximation to the Poisson statistics. We restrict our attention to small time intervals $t \in [t_0, t_0 + \Delta t]$ containing the resonance time $t_r$, when the number of magnons at time $t_0$ is $N_0$. So long as the number of magnons lost is small compared to $N_0$, we may characterize the Gaussian distribution of magnon loss by a mean $\Gamma \Delta t N_0$, which is also equal to the variance.

Subject to these assumptions and approximations, the random variable $\Delta_{i}$, the number of magnons lost between times $t_{i-1}$ and $t_{i}$, has statistics
\begin{gather}
\label{eq:avgmagnonloss}
\mathbf{E}[\Delta_{i} ] = \Gamma N_0 (t_{i} - t_{i-1}),\\
\mathbf{Cov}[\Delta_{i}, \Delta_{j}] = \delta_{ij} \Gamma N_0 (t_{i}-t_{i-1}),
\end{gather}
where $\delta_{ij}$ is the Kronecker delta. Next, we can write the number of magnons $N_i$ at $t_i$ as 
\begin{equation}
    N_{i} = N_0 - \sum^{i-1}_{j=0} \Delta_j.
\end{equation}
As the sum of Gaussian variates, $N_{i}$ will be Gaussian-distributed, with statistics
\begin{gather}
    \mathbf{E}[N_i] = N_0[1-(t_i-t_0)\Gamma], \\
    \mathbf{Cov}[N_i,N_j] = N_0 [\mathrm{min}(t_i,t_j) -t_0] \Gamma.
\end{gather}
Assuming the $t_i$ are uniformly spaced by $\Delta t$, we have
\begin{gather}
    \mathbf{E}[N_i] =N_0 (1 - i \Delta t \Gamma), \\
    \mathbf{Cov}[N_i,N_j] = \mathrm{min}(i, j)  N_0 \Gamma \Delta t  .
\end{gather}

To relate the statistics of the magnon number to the statistics of the domain wall location (in absence of an axion wind), we use that the total magnon number in the system as a function of the domain wall position $z$ is given by 
\begin{equation}
    N = \frac{5}{4} \frac{\chi B_0(1 + \alpha z)}{\gamma} V_\mathrm{HPD},
    \label{eq:TotalMagnonNumber}
\end{equation}
where $V_\mathrm{HPD} = A z$. Defining $z(N)$ as the inverse of Eq.~(\ref{eq:TotalMagnonNumber}), for our small time interval around the resonance, we have 
\begin{equation}
    z(t_i) \approx  z(t_0) + \frac{dz}{dN} N_i,
\end{equation}
where $dz/dN$ as evaluated at $N = N_0$. Since this is an affine transformation, we have
\begin{gather}
    \mathbf{E}[z_0(t_i)] = z(t_0) + N_0 (1 - i \Gamma \Delta t )  \frac{dz}{dN} \label{eq:NullMean}, \\
    \mathbf{Cov}[z_0(t_i), z_0(t_j)] =  \mathrm{min}(i, j) N_0 \Gamma \Delta t \left(\frac{dz}{dN}\right)^2.
    \label{eq:NullCov}
\end{gather}
Note that we have re-introduced subscript zeros indicating that this is the covariance for the domain wall motion in absence of an axion. It remains to modify this expression in the context of an axion-induced resonance.

\subsection{Domain Wall Motion from a Stochastic Axion Wind}
We now consider how the presence of an axion field modifies the statistics of the domain wall motion. Recalling that $z_a$ was defined as the dimensionless domain wall shift relative to $z_0$ induced by the axion, we must now calculate the mean and covariance of the quantity $z(t) = z_0(t)(1 + z_a(t))$. First, for the mean, we have
\begin{equation}
\begin{split}
\mathbf{E}[z(t)] = \mathbf{E}[z_0(t)] + \mathbf{E}[z_0(t) z_a(t)].
\end{split}
\end{equation}
From Eq.~(\ref{eq:zEoM}), $\dot z_a$ is linear in $\mathbf{B}_a$, which has a magnitude determined by the uniformly distributed axion phases and is independent of all other variates. Computing any expectation value linear in $z_a$ is equivalent to computing the expectation value of a weighted sum of quantities linear in $\dot z_a$, which must vanish when taking the expectation value over the axion phase. Thus, we find even in the presence of an axion gradient, the mean domain wall position is equal to its background value,
\begin{equation}
    \mathbf{E}[z(t_i)] =  \mathbf{E}[z_0(t_i)],
\end{equation}
which is explicitly evaluated in Eq.~(\ref{eq:NullMean}).

Now we proceed to the more complicated covariance term
\begin{equation}
\begin{split}
\mathbf{Cov}[z(t_i) z(t_j)] &= \mathbf{Cov}[z_0(t_i)z_0(t_j)] \\
& + \mathbf{Cov}[z_0(t_i), z_0(t_j)z_a(t_j)] \\
& + \mathbf{Cov}[z_0(t_i)z_a(t_i), z_0(t_j)] \\
&+ \mathbf{Cov}[z_0(t_i)z_a(t_i), z_0(t_j)z_a(t_j)],
\label{Eq:TotalCov}
\end{split}
\end{equation}
which we consider term-by-term. The first term is the background covariance calculated in Eq.~(\ref{eq:NullCov}). The second and third terms can be expanded into products of expectation values which are linear in $z_a$, and thus vanish by the arguments above. To deal with the fourth term, which is quadratic in both $z_0$ and $z_a$, we write out this covariance explicitly in terms of expectation values:
\begin{equation}
\begin{split}
\mathbf{Cov}[z_0(t_i) z_a(t_i), z_0(t_j) z_a(t_j)] &= \mathbf{E}[z_0(t_i) z_0(t_j) z_a(t_i) z_a(t_j)] \\
& - \mathbf{E}[z_0(t_i)z_a(t_i)] \mathbf{E}[z_0(t_i)z_a(t_j)] \\
&=  \mathbf{E}[z_0(t_i) z_0(t_j) z_a(t_i) z_a(t_j)].
\end{split}
\end{equation}
The last line follows because quantities which are linear in $z_a$ vanish. Defining $\delta z_0(t_i) \equiv z_0(t_i) - \mathbf{E}[z_0(t_i)]$, we may write
\begin{widetext}
\begin{equation}
\begin{split}
\mathbf{Cov}[z_0(t_i) z_a(t_i), z_0(t_j) z_a(t_j)]  &x=  \mathbf{E}[z_0(t_i) z_0(t_j) z_a(t_i) z_a(t_j)]\\
&= \mathbf{E}\bigg[\bigg(\mathbf{E}[z_0(t_i)] +\delta z_0(t_i) \bigg) \bigg(\mathbf{E}[z_0(t_j)] +\delta z_0(t_j) \bigg) z_a(t_i) z_a(t_j)\bigg]\\
&= \mathbf{E}[z_0(t_i)]\mathbf{E}[z_0(t_j)] \mathbf{E}[z_a(t_i) z_a(t_j)] + \mathbf{E}[z_0(t_i)] \mathbf{E}[\delta z_0(t_j) z_a(t_i) z_a(t_j)] \\
&\ \ \ + \mathbf{E}[z_0(t_j)] \mathbf{E}[\delta z_0(t_i) z_a(t_i) z_a(t_j)] +  \mathbf{E}[\delta z_0(t_i)  \delta z_0(t_j) z_a(t_i) z_a(t_j)].
\label{eq:FullCov}
\end{split}
\end{equation}
\end{widetext}
Over a small time interval $\delta T$ around the resonance, $\delta z_0 / z_0 \sim \delta T / T_1 \ll 1$, and so neglecting terms which are proportional to $\delta T / T_1$, we have
\begin{equation}
\begin{split}
    \mathbf{Cov}[&z_0(t_i) z_a(t_i),z_0(t_j) z_a(t_j)]  \\
    &\approx \mathbf{E}[z_0(t_i)]\mathbf{E}[z_0(t_j)] \mathbf{E}[z_a(t_i) z_a(t_j)] \\
    &=  \mathbf{E}[z_0(t_i)]\mathbf{E}[z_0(t_j)] \mathbf{Cov}[z_a(t_i), z_a(t_j)]
\label{eq:ReducedCov}
\end{split}
\end{equation}
at leading order in $g_{aNN}$.

Collecting the above results, we obtain
\begin{equation}
\begin{split}
\mathbf{Cov}[z(t_i), &z(t_j)] = \mathbf{Cov}[z_0(t_i), z_0(t_j)] \\
& + \mathbf{E}[z_0(t_i)]\mathbf{E}[z_0(t_j)] \mathbf{Cov}[z_a(t_i), z_a(t_j)].
\label{eq:TotalCov}
\end{split}
\end{equation}
The covariance has nicely decomposed into the sum of terms which do and do not depend on the axion gradient field, which will facilitate a computation of the likelihood function for our signal.

\subsection{Stochastic Evolution of the Precession Frequency}

Finally, we must relate the mean and covariance of the domain wall position to the mean and covariance of the precession frequency $\omega_L$. Since $\omega_{L,i} \equiv \dot{\theta}(t_i)$ is given by
\begin{equation}
    \omega_{L,i} = \gamma B_0 [1 + \alpha z(t_i)],
\end{equation} 
its mean $\boldsymbol{\mu}$ and covariance $\boldsymbol{\Sigma}$ are given by
\begin{gather}
    \boldsymbol{\mu}_i \equiv \mathbf{E}[\omega_{L,i}] = \gamma B_0 \left[1 + \alpha \mathbf{E}[z_i] \right], \\
    \boldsymbol{\Sigma}_{ij} = \boldsymbol{B}_{ij} + \boldsymbol{S}_{ij}.
\end{gather}
where we have split the covariance into a background term $\boldsymbol{B}$ (not to be confused with the external magnetic field $\mathbf{B}$) and a signal covariance $\boldsymbol{S}$:
\begin{gather}
\boldsymbol{B}_{ij} = (\alpha \gamma B_0)^2 \mathbf{Cov}[z_0(t_i), z_0(t_j)], \\ 
\boldsymbol{S}_{ij} = (\alpha \gamma B_0)^2 \mathbf{E}[z_0(t_i)]\mathbf{E}[z_0(t_j)] \mathbf{Cov}[z_a(t_i), z_a(t_j)].
\end{gather}

There exist some remaining subtleties that we have not fully addressed. First, we have only calculated the first two moments of $\omega_{L,i}$. These moments fully characterize Gaussian distributions, but even for Gaussian $z_0$ and $z_a$, the product $z_0 z_a$ will generically be non-Gaussian after sufficiently long times, motivating careful choice of duration of windows during which axion resonances are searched for in data. Similarly, the only computationally tractable way to calculate $\langle z_a(t) z_a(t') \rangle$ is to fix $z_0(t) = \mathbf{E}[z_0(t)]$, which we expect to be accurate to at $\mathcal{O}(\delta z_0 / z_0) \sim \mathcal{O}(\delta T / T_1)$. Though we have been somewhat schematic here, in App.~\ref{App:SDE}, we demonstrate with numerical tests that our approximations are good ones. Also, note that these caveats regarding Gaussianity and the order of accuracy in $\delta z_0 / z_0$ apply only to the contribution of the axion-induced domain wall motion, and so in the small signal limit, they are further suppressed relative to the expected domain wall motion in absence of an axion by the small axion coupling.

As a final comment, we point out that the covariance we have developed describes only the stochasticity of the HPD system and does not account for any variety of measurement error, which will act as an additional source of nonzero covariance. 

\subsection{Likelihood for HPD Measurements}
\label{sec:Statistics}

Since our data approximately follows a multivariate Gaussian distribution, it has likelihood function
\begin{equation}
    \mathcal{L}(\mathbf d, \mathcal{M}, \boldsymbol{\theta}) = \frac{\exp\left[-\frac{1}{2} (\mathbf{d}-\boldsymbol{\mu})^T \mathbf{\Sigma}^{-1}(\boldsymbol{\theta}) (\mathbf{d} -\boldsymbol{\mu}) \right]}{\sqrt{(2\pi)^{N} |\mathbf{\Sigma}(\boldsymbol{\theta})|}},
\end{equation}
where $\boldsymbol{\mu}$ and $\mathbf{\Sigma}$ are the mean and covariance, respectively, for a model $\mathcal{M}$ parameterized by $\boldsymbol{\theta}$, $\mathbf{d}$ is the observed data consisting of $N$ datapoints, and $|\boldsymbol{\Sigma}| \equiv {\rm det} (\boldsymbol{\Sigma})$. From this likelihood, we define a test statistic $\Theta$ for discovery in terms of the signal parameter $S$,
\begin{equation}
    \Theta(S) = 2 \left[\ln \mathcal{L}(d | \hat{\boldsymbol{\theta}}_b, S)-\ln \mathcal{L}(d | \hat{\boldsymbol{\theta}}_b, S=0) \right],
\end{equation}
where $\boldsymbol{\theta}_b$ are the nuisance parameters, which may include $T_1$, $\alpha$, and any other parameters relevant for fully characterizing the system and the measurement (such as clock noise, discussed in Sec.~\ref{sec:Sensitivity} below). We note that incorporating these and other relevant nuisance parameters enables a characterization of the system \textit{in situ} and in the specific small interval in time relevant for a given axion signal. However, for simplicity in our sensitivity estimates, we assume that all nuisance parameters may be determined with perfect accuracy so that $\hat{\boldsymbol{\theta}}_b$ denotes the maximum likelihood estimators of those nuisance parameters under the null hypothesis.

Evaluating this likelihood, we have 
\begin{equation}
    \Theta(S) = (\mathbf{d}-\boldsymbol{\mu})^T \left[\boldsymbol{B}^{-1} - \mathbf{\Sigma}^{-1}\right] (\mathbf{d}-\boldsymbol{\mu}) - \ln \left[\frac{|\mathbf{\Sigma}|}{|\boldsymbol{B}|}\right],
\end{equation}
where $\boldsymbol{B}$ is the background-only covariance and $\boldsymbol{\Sigma}$ is the covariance including the signal contribution. Next, following the procedure of \cite{Cowan:2010js}, we  evaluate the Asimov expected test statistic under the assumption of some true signal strength $S^t$, as follows:
\begin{equation}
\begin{split}
    \Theta &= \langle (\mathbf{d}-\boldsymbol{\mu})^T \left[\boldsymbol{B}^{-1} - \mathbf{\Sigma}^{-1}\right] (\mathbf{d}-\boldsymbol{\mu}) \rangle  - \ln \left[\frac{|\mathbf{\Sigma}|}{|\boldsymbol{B}|}\right] \\
           &= \mathrm{Tr} \left(\langle (\mathbf{d}-\boldsymbol{\mu}) (\mathbf{d}-\boldsymbol{\mu})^T \rangle  \left[\boldsymbol{B}^{-1} - \mathbf{\Sigma}^{-1}\right] \right)  - \ln \left[\frac{|\mathbf{\Sigma}|}{|\boldsymbol{B}|}\right] \\
           &= \mathrm{Tr} \left(\mathbf{\Sigma}^t \left[\boldsymbol{B}^{-1} - \mathbf{\Sigma}^{-1}\right] \right) - \ln \left[\frac{|\mathbf{\Sigma}|}{|\boldsymbol{B}|}\right],
\end{split}
\end{equation}
where $\mathbf{\Sigma}^t$ is the true covariance of the data.  Fixing the background at its true value, we can write
\begin{equation}
    \mathbf{\Sigma}^t = \boldsymbol{S}^t + \boldsymbol{B},
\end{equation}
and expand to quadratic order in the signal strength to obtain
\begin{equation}
    \Theta \approx \mathrm{Tr} \left[ \left(\boldsymbol{S}^t - \frac{1}{2} \boldsymbol{S} \right) \boldsymbol{B}^{-1} \boldsymbol{S} \boldsymbol{B}^{-1} \right],
    \label{eq:TS}
\end{equation}
where we have used the standard identity $\ln |{\bf M}|={\rm Tr}(\ln {\bf M})$. Note that the test statistic in Eq.~(\ref{eq:TS}) takes the same form as in the ``axion interferometry'' search of Ref.~\cite{Foster:2020fln}. This is because the formalism of Ref.~\cite{Foster:2020fln} holds for any positive-definite background covariance $\boldsymbol{B}$, which indeed is a requirement for a physically-reasonable covariance matrix.

The signal covariance can be written with canonical normalization $\boldsymbol{S} \rightarrow A \boldsymbol{S}$, for $A = g_{aNN}^2$ so the Asimov expected sensitivity given the true value of the coupling $A^t$ becomes
\begin{equation}
    \Theta(A |A^t ) = A\left(A^t - \frac{1}{2}A\right)\mathrm{Tr} \bigg[ (\boldsymbol{S} \boldsymbol{B}^{-1})^2 \bigg].
\end{equation}
This is maximized for $A = A^t$, which shows that our likelihood is an unbiased estimator. Evaluating the expected sensitivity to $g_{aNN}$ under the null hypothesis ($g_{aNN}^t = 0)$ using the Fisher information \cite{Rao, Cramer}, we obtain
\begin{equation}
    \sigma_{A}^{-2} = \frac{1}{2}\mathrm{Tr} \left[ (\boldsymbol{S} \boldsymbol{B}^{-1})^2 \right].
    \label{eq:FisherInformation}
\end{equation}
From here, we are able to compute the expected sensitivity for arbitrary data collection and analysis schemes. For instance, the expected 95$^\mathrm{th}$ percentile upper limit on $g_{aNN}$ is defined by
\begin{equation}
    g_{aNN}^{95} = \left[ \Phi^{-1}(0.95) \sigma_A\right]^{1/2},
    \label{eq:expectedSensitivity}
\end{equation}
where $\Phi^{-1}(x)$ is the inverse of the standard normal distribution.

We also note that the HPD relaxation time $T_1$ is typically larger than the axion coherence time in Eq.~(\ref{eq:CoherenceTime}), and so under these conditions, any repeated measurements will probe effectively uncorrelated gradient field realizations. As a result, $N$ repeated measurements will add linearly in the test statistic of Eq.~(\ref{eq:TS}), leading to a $N^{1/4}$ enhancement of the sensitivity of Eq.~(\ref{eq:expectedSensitivity}). Hence, we have $g_{aNN}^{95}\propto t_{\rm int}^{-1/4}$ where $t_{\rm int}$ is the total integration time.\footnote{This behavior is consistent with the analysis of Ref.~\cite{Dror:2022xpi}, since in the HPD there is only one relaxation timescale $T_1$, and over multiple runs with $t_{\rm int} \gg T_1$, the overall scaling is $t_{\rm int}^{-1/4}$.}

In all subsequent examples and calculations, we take the local DM velocity distribution to be given by Standard Halo Model ansatz of
\begin{equation}
f(\mathbf{v}) = \frac{1}{[2 \pi \sigma_v^2]^{3/2}} \exp\bigg[-\frac{|\mathbf{v} - \mathbf{v}_\mathrm{obs}|^2}{2\sigma_v^2}\bigg]
\end{equation}
where $\mathrm{v}_\mathrm{obs}$ is the lab velocity in the halo frame and $\sigma_v$ is the velocity dispersion of $155\, \mathrm{km/s}$ \cite{Lisanti:2016jxe}. We initially take the lab-frame velocity to be $\mathrm{v}_\mathrm{obs} = 230\, \mathrm{km/s}$ in the $\hat{\mathbf{x}}$ direction, corresponding to the solar velocity with respect to the galactic halo, but we relax this assumption when considering temporal modulation effects in Sec.~\ref{sec:Modulation}.

\section{HPD Measurement and Projected Sensitivity}
\label{sec:Sensitivity}

An important practical consequence of our analysis above is that measuring the evolution of the precession frequency with fine time resolution is particularly important for maximizing the sensitivity of HPD-based searches for axions. This is because although a resonance may induce an overall shift in $\omega_L$ with respect to the expectation (as was considered to be the signal in Ref.~\cite{Gao:2022nuq}), the same effect is produced by stochastic magnon loss. The variance in the magnon loss over a finite time interval grows linearly with the duration of that interval, as illustrated in the grey band of Fig.~\ref{fig:SigCovExample}, so our ability to identify a resonance at some time $\Delta t$ after it occurs degrades as $\sqrt{\Delta t}$. Moreover, as we have seen in Fig.~\ref{fig:SigCovExample}, ringing features associated with beat frequency effects appear in the axion-induced shift in $\omega_L$ that are not expected if the precession rate evolution is governed solely by stochastic magnon loss.  Good time resolution in measurements of $\omega_L$ enables identification of these features and thus allows for improved sensitivities. We note that recent work in \cite{Dror:2022xpi} revealed previously unappreciated aspects of the time-dependent scaling of the axion coupling sensitivity of spin precession experiments. In our work, these details are accounted for by construction by the explicit time-integration of Eq.~(\ref{eq:zEoM}). In this section, we describe a concrete proof-of-principle measurement and readout scheme which accounts for imprecision in both the frequency measurement and the reference clock, and show how the interplay of these noise sources with the irreducible stochastic noise informs our choice of measurement cadence and scanning strategy.

\subsection{Frequency Measurement with Optimal Quantum Control}

Since the axion signal we are seeking appears only in the precession frequency and not the amplitude, our goal is a frequency measurement scheme which is as insensitive as possible to amplitude noise but maximally sensitive to frequency shifts. This implies that extracting the precession phase from e.g.\ measurements of $M_x$ and $M_y$ with orthogonal pickup loops will be polluted by amplitude noise from the pickup loops. Furthermore, simply performing a Fourier transform on the resulting signal results in a frequency resolution $\delta \omega/\omega$ which scales as $1/T$, where $T$ is the total measurement time.

Both of these situations may be vastly improved with the techniques of quantum metrology~\cite{pang2017optimal}. The precessing HPD magnetization may couple to a quantum system (such as a qubit) as a time-dependent Hamiltonian. The Heisenberg bound implies that errors on estimates of a parameter from a time-\emph{independent} Hamiltonian scales at best like $1/T$, but for a time-\emph{dependent} Hamiltonian, one may carefully choose an additional time-dependent control Hamiltonian $H_c(t)$ to manipulate the system and achieve in principle arbitrary scaling with $T$, limited only by the specific time dependence of the Hamiltonian to be estimated. In the particular case of a frequency measurement, Ref.~\cite{pang2017optimal} shows that with optimal quantum control (i.e.\ maximizing the quantum Fisher information), the measurement of the frequency of a rotating magnetic field using a single qubit has errors that scale like $1/T^2$. Furthermore, Ref.~\cite{naghiloo2017achieving} demonstrated this scaling experimentally in a closely-related setup where an external Hamiltonian modulated the level spacing in a qubit, and the task was to estimate the modulation frequency $\omega$. While these scalings only hold up to the qubit coherence time $T_q$, Refs.~\cite{boss2017quantum,schmitt2017submillihertz} derived, and confirmed experimentally, that a frequency error scaling of $1/\sqrt{T_q T^3}$ may be achieved for $T > T_q$ with a heterodyne readout scheme referenced to a stable external clock.

Since our goal is actually to measure a frequency \emph{drift} $\dot{\omega}$, it turns out we can achieve even more favorable scaling with measurement time. Our readout scheme is as follows. The transverse HPD magnetization is sensed with a SQUID oriented in the $xy$-plane. Rather than reading out the SQUID directly and using it as a magnetometer, we imagine coupling the SQUID to $N_q$ transmon qubits. The time-varying flux through the SQUID acts as a time-dependent modulation of the qubit level splitting, $H_{\dot{\omega}}(t) = A \sin(\omega_L t + \dot{\omega} t^2/2) \sigma_z/2$, exactly analogous to Ref.~\cite{naghiloo2017achieving}. The amplitude $A$ of the level splitting modulation is proportional to the amplitude $\widetilde{M}_0$ of the HPD magnetization. With $B_0 = 0.5 \ {\rm T}$, $\widetilde{M}_0 \simeq 50 \ {\rm nT}$, and the size of the modulation in frequency units is on the order of $A \simeq 2\pi \times 10 \ {\rm MHz}$ for a qubit with typical frequency 6 GHz~\cite{MurchPrivate}. The control Hamiltonian $H_c(t)$ is applied in order to maximize the quantum Fisher information; see App.~\ref{sec:OptimalControlMeasurement} for details.

We then estimate the Hamiltonian parameter $\dot{\omega}$ by reading out the state of the qubits at a measurement cadence of time interval $\Delta t$, where the start and end of each measurement interval are referenced to an external clock. Following the derivation of Ref.~\cite{pang2017optimal}, the error on $\dot{\omega}$ scales as
\begin{equation}\label{eq:readouterror_omegadot1}
    \delta \dot{\omega} = \frac{3\pi}{A (\Delta t)^3 \sqrt{4N_q}} \qquad (\Delta t \leq T_q),
\end{equation}
where the scaling with $1/\sqrt{N_q}$ is the standard spin projection noise from $N_q$ independent qubits. The $T^3$ scaling for a measurement of $\dot{\omega}$ compared to the $T^2$ scaling for a measurement of $\omega$ arises simply from an additional factor of $t$ in the desired Hamiltonian $\partial_{\dot{\omega}} H_{\dot{\omega}}$. Following the derivation of Ref.~\cite{schmitt2017submillihertz}, for measurement cadences exceeding $T_q$ we instead read out the qubit states at intervals of $T_q$, record the results to disk with timestamps given by the external clock, and obtain
\begin{equation}\label{eq:readouterror_omegadot2}
    \delta \dot{\omega} = \frac{\sqrt{5}\pi }{A (\Delta t)^{5/2} T_q^{1/2}\sqrt{4N_q}} \qquad (\Delta t \geq T_q).
\end{equation}
For details, see App.~\ref{sec:OptimalControlMeasurement}. Once we have measured $\dot{\omega}$, we then define our measured frequency stepwise over each measurement cadence between $t_0$ and $t_0 + \Delta t$ as a linear approximation
\begin{equation}
    \omega(t) = \omega(t_0) + \dot{\omega} \times (t - t_0).
\end{equation}
From this point on, we reference to the rate $1/\Delta t$ at which we make optimally controlled measurements of $\dot \omega$ as the measurement cadence $f$. 

As we discuss in App.~\ref{sec:OptimalControlMeasurement}, both $\omega$ and $\dot \omega$ are measured in the optimal control scheme by querying the same set of qubit states. As a result, there exists some degeneracy in measuring these quantities, and measurement of $\dot \omega$ at the precision we project requires a previous measurement of $\omega$ at precision such that $\delta \omega / \Delta t \ll \delta \dot{\omega}$ where $\delta \omega$ is the uncertainty on $\omega$ made in previous measurements, $\Delta t$ is the time interval of the measurement cadence, and $\delta \dot{\omega}$ is the desired precision of the $\dot \omega$ measurement. In practice, the time necessary to make a sufficiently precise measurement of $\omega$ is small compared to the HPD relaxation times of interest in this work. As a result, only a small fraction of the total integration time must be dedicated to determining $\omega$, and this requirement has a negligible impact on projected sensitivities.

Finally, since our measurement of $\omega$ relies on the timing of our measurement cadence, imprecision in the external clock acts as a noise floor that must be considered in tandem with the measurement noise on $\dot{\omega}$. We account for this imprecision by adding a diagonal covariance 
\begin{equation}
    \boldsymbol{B}^\mathrm{clock}_{ij} = B_\mathrm{clock} \times \left(\frac{1\,\mathrm{s}}{\Delta t} \right) \omega_{L,i} \omega_{L,j} \delta_{ij}.
\end{equation}
Here $B_\mathrm{clock}$ is the relative clock variance over a $1\,\mathrm{s}$ interval, and we have assumed the clock to be performing at the white noise limit so that the variance is inversely related to the integration time. We assume a clock precision $B_\mathrm{clock} = 3.6 \times 10^{-31}$ associated with a cryogenic sapphire microwave oscillator, though even higher-performing alternatives may be possible \cite{doi:10.1126/science.abb2473}.\footnote{Above $1\,\mathrm{s}$, the sapphire oscillator precision scales less rapidly than $(\Delta t)^{-1/2}$. Assuming the precision scales like white noise for integration times less than $1\,\mathrm{s}$ therefore represents a conservative assumption.}

\subsection{Sensitivity as a Function of Measurement Cadence}
\label{sec:ClockSensitivity}
As a benchmark scenario, we consider an HPD system of height $h = 10\,\mathrm{cm}$ and total volume $100\,\mathrm{cm}^3$ with relaxation time $T_1 = 1000\,\mathrm{s}$, which is longer than the $T_1$ which has so far been measured in the HPD but was argued in Ref.~\cite{Gao:2022nuq} to be a reasonable goal for an experimental program. We take $B_0 = 0.05\,$T and $\alpha = 1\,\mathrm{cm}^{-1}$. Therefore, 
$N_0\sim 10^{20}$. The background covariance from the stochastic magnon loss is typically $\frac{\boldsymbol{B}_{ij}}{\omega_{L,i}\omega_{L,j}}\sim \frac{\Delta t}{4 T_1 N_0}\sim 10^{-24}\frac{\Delta t}{1\,{\rm s}}$.

We take the domain wall to be initially located at $z(t_0) = h = 10\,\mathrm{cm}$ at $t_0 = 0\,\mathrm{s}$ and consider an axion mass $m_a = 73\,\mathrm{neV}$, corresponding to $\omega_L/(2\pi) = 17.6 \, {\rm MHz}$ at the time of resonance.

\begin{figure}[t!]  
\hspace{0pt}
\vspace{-0.2in}
\begin{center}
\includegraphics[width=0.49\textwidth]{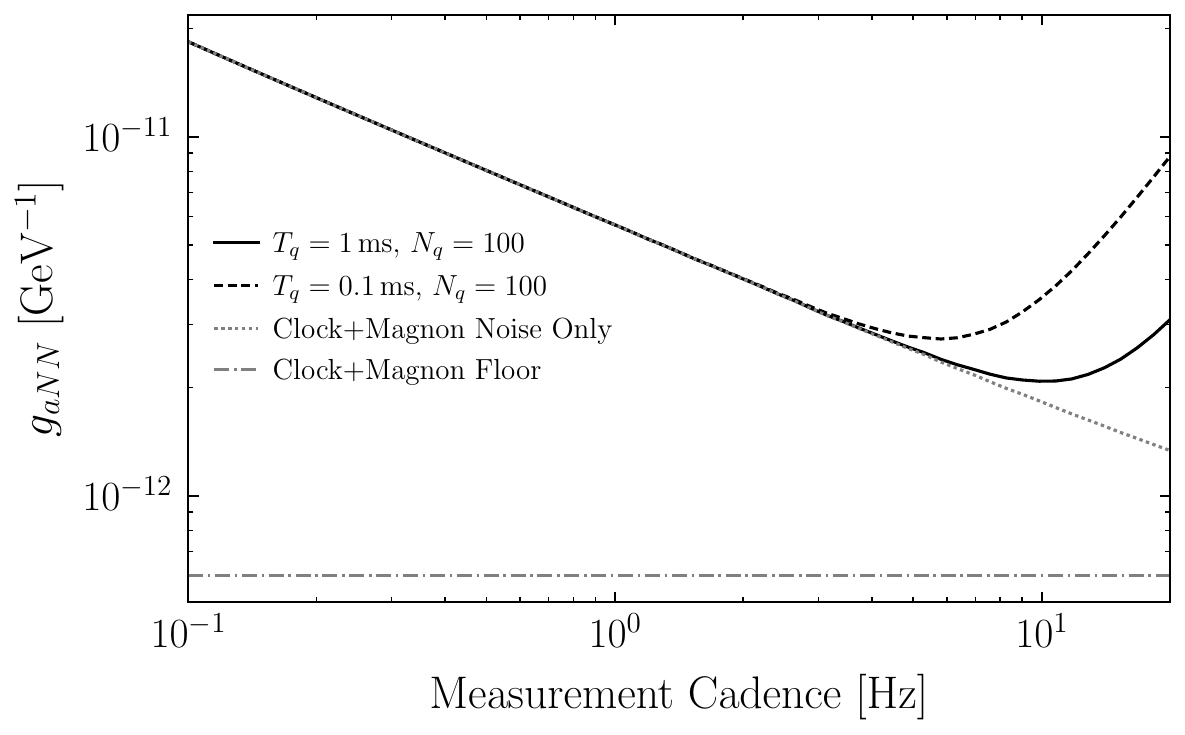}
\caption{The axion-coupling sensitivity of 1 month of HPD measurement for an axion of mass $m_a = 73\,\mathrm{neV}$ as a function of measurement cadence. Two different qubit measurement scenarios for $N_q = 100$ and $N_q = 1$ are shown in solid black and dashed black, respectively. To illustrate the importance of measurement imprecision, we also show the sensitivity as a function of cadence in the absence of measurement error in dotted grey. In this idealized case, the coupling sensitivity would continue to improve until it saturates at $f \approx 1\,\mathrm{kHz}$ at a floor indicated by the grey dot-dashed line. For details, see the main text in  Sec.~\ref{sec:ClockSensitivity}.}
\label{fig:FreqProj}
\end{center}
\end{figure}

We can determine the limit-setting power of an HPD measurement as a function of our frequency-measuring strategy, accounting for the joint effects of magnon stochasticity, clock noise, and measurement imprecision. Parametrically, the clock noise scales with the measurement cadence as $\sqrt{f}$ while the magnon loss noise scales as $1/\sqrt{f}$. These are relatively slow scalings with $f$, and by contrast, the measurement imprecision scales as either $f^6$ or $f^{5/2}$ depending on the choice of cadence with respect to the qubit coherence time. Faster cadences allow for better resolution of the resonance, effectively reducing the magnon noise relevant for searching for a signal, but will suffer from considerably larger clock and measurement imprecision. 

Fig.~\ref{fig:FreqProj} shows the sensitivity as a function of measurement cadence for three illustrative scenarios. In an optimistic scenario (solid curve), we project a measurement performed using $N_q = 100$ qubits, each with a coherence time of $T_q = 1.0 \ \mathrm{ms}$. This coherence time in superconducting qubits is presently achievable~\cite{wang2022towards}, with theoretical upper bounds as large as $T_q \approx 3\,\mathrm{ms}$~\cite{Read:2022cpk}. In a more conservative scenario (dashed curve), we assume a smaller coherence time $T_q = 0.1\,\mathrm{ms}$, but with the same $N_q =100$.\footnote{Since the noise scales identically with $T_q$ and $N_q$, the solid curve may also be realized with $T_q = 0.1\,\mathrm{ms}$ and $N_q =1000$.} Finally, the dotted curve shows the unphysical scenario in which the measurement imprecision is vanishing and only the clock imprecision and magnon loss stochasticity contribute to the noise. In the optimistic scenario, the optimal measurement cadence is $f \approx 5\,\mathrm{Hz}$ due to the rapid scaling of measurement imprecision with $f$. This cadence is too slow to resolve either the axion resonance, which lasts about 50 ms (see Fig.~\ref{fig:SigCovExample}), or the beat frequencies after the resonance, resulting in a loss of sensitivity. To resolve the resonance at a cadence of 80 Hz, one would need $N_q = 10^6$ with $T_q = 1 \ {\rm ms}$, and to reach the clock+magnon noise floor, one would need $N_q = 10^8$, which is (needless to say) unrealistic in the near future. That said, additional improvements to the sensitivity may be gained via an optimization of the qubit coupling parameter $A$, though a careful analysis accounting for backaction noise (which would affect $T_q$) is necessary.

\subsection{Sensitivity as a Function of B-Field Gradient}
\label{sec:GradSensitivity}

\begin{figure}[t!]  
\hspace{0pt}
\vspace{-0.2in}
\begin{center}
\includegraphics[width=0.49\textwidth]{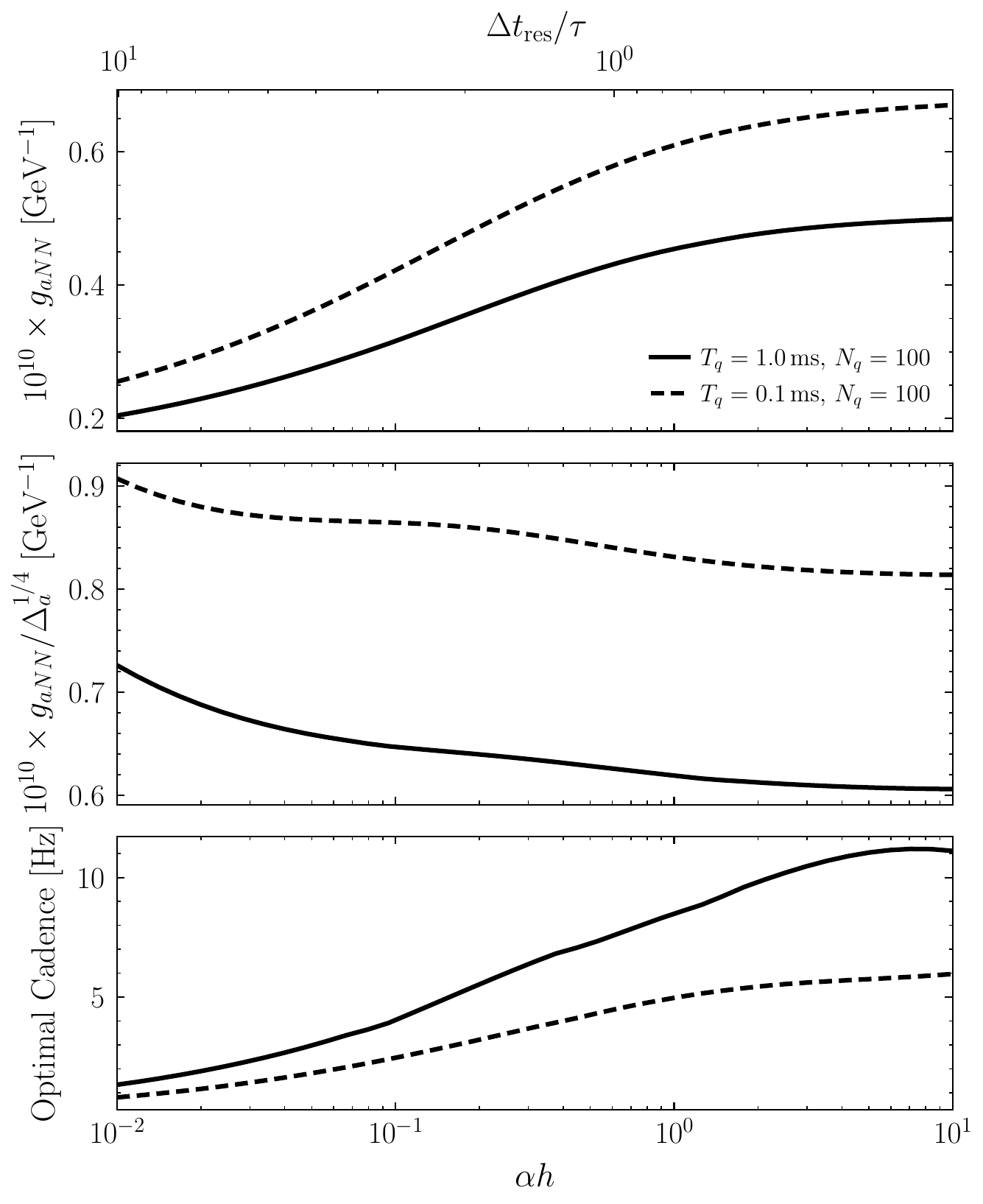}
\caption{(\textit{Above}) The coupling sensitivity assuming a single data collection for time $T_1$ with an axion of mass $m_a = \mathrm{73}\,\mathrm{neV}$, as a function of the time on resonance in the scan relative to the axion coherence time $\tau$. We present this coupling sensitivity for our two qubit scenarios, see text for details. The time on resonance for a fixed $T_1$ is set by the $B$-field gradient $\alpha$, and so all three panels share an $x$-axis. (\textit{Middle}) The figure of merit for axion coupling sensitivity (see Sec.~\ref{sec:GradSensitivity} for details), as a function of the external $B$-field gradient for our two qubit scenarios. Decreasing values indicate improved integrated sensitivity due to the tradeoff between scan time and scan range. For all scenarios, scan sensitivity is generally saturated by $\alpha h \approx 1$. (\textit{Below}) The optimal measurement cadence as a function of $\alpha h$, which we find is not a strong function of the external $B$-field gradient.}
\label{fig:GradProj}
\end{center}
\end{figure}

Axion DM experiments all suffer from the unfortunate fact that the true axion mass is unknown. When aiming for an integrated sensitivity across a range of possible axion masses, the choice of $\alpha$ must be made carefully. For large $\alpha$, a much larger range of precession frequencies are realized, providing sensitivity at a larger range of masses, but generally at worse sensitivity for a given mass than if $\alpha$ were small. On the other hand, smaller values of $\alpha$ will require more scans to cover an identical range of masses for a fixed total data-taking time. Defining $\Delta_a = 2(\dot\theta_\mathrm{max} - \dot\theta_\mathrm{min}) / (\dot\theta_\mathrm{max} + \dot\theta_\mathrm{min})$ as the relative size of the mass interval reached by a single collection at a given $\alpha$, a scanning strategy can be roughly optimized by minimizing the quantity $g_{aNN}^{95}(\alpha) / (\Delta_a(\alpha))^{1/4}$.

To calculate the typical axion coupling sensitivity as a function of $\alpha$, we maintain our previous set of benchmark parameters from Sec.~\ref{sec:ClockSensitivity} when possible. We take $h = 10\,\mathrm{cm}$, total volume $100 \,\mathrm{cm}^3$, and relaxation time $T_1 = 1000\,\mathrm{s}$.  As before, we consider two measurement scenarios, $T_q= 1.\,\mathrm{ms}$ and $T_q = 0.1\,\mathrm{ms}$, each with $N_q = 100$. For each choice of $\alpha$ and $N_q$ we independently optimize the measurement cadence in computing the sensitivity. We evaluate the coupling sensitivity for a mass $m_a = 73\,\mathrm{neV}$ for a range of $\alpha$, choosing $B_0$ self-consistently so that the resonance occurs when the system is at a height of $9.99\,\mathrm{cm}$ and the number of magnons in the system is the same at the time of resonance for all choices of $\alpha$.

We first consider the sensitivity at a fixed axion mass, shown in Fig.~\ref{fig:GradProj} (top). The $B$-field gradient is inversely proportional to the time on resonance, $\Delta t_{\rm res}$, which is defined as the time interval over which $\omega_L$ has drifted by a fraction $10^{-6}$ corresponding to the axion bandwidth. For our parameters, $\alpha h \simeq 0.1$ corresponds to $\Delta t_{\rm res} \simeq \tau$, which was the situation studied in Ref.~\cite{Gao:2022nuq}. We see that an optimized measurement achieves the best sensitivity to a single mass with the longest time on resonance, but that the sensitivity changes only by a factor of $3$ going from $\alpha h = 10^{-2}$ to $\alpha h = 10^{1}$. This behavior as a function of resonance timescale is significantly different than the scaling obtained in Ref.~\cite{Dror:2022xpi} for an experiment such as CASPEr-Wind which is limited by amplitude noise and for which the time on resonance is simply given by the total measurement time. In our setup, the combined effects of clock noise, magnon noise and measurement noise intertwine time-domain and frequency-domain phenomena, making the scaling of the $g_{aNN}$ limits with $\Delta t_{\rm res}$ less straightforward.

Next, we consider the figure of merit $g_{aNN}/\Delta_a^{1/4}$ in the middle panel of Fig.~\ref{fig:GradProj}. We generally find that the figure of merit is saturated by $\alpha h \approx 1$, motivating somewhat larger field gradients than considered in Ref.~\cite{Gao:2022nuq}. We find that although sensitivity to an individual mass is improved by increasing time on resonance, that the best integrated sensitivity of a scan over a range of masses is achieved by $\alpha h\approx 1$. We caution, though, that by calculating only a single representative axion coupling sensitivity, we have neglected the dependence of coupling sensitivity on the precise height of the domain wall at the time of the resonance; an optimization of a full-fledged experimental procedure would require a more detailed treatment. Finally, we also show the optimal measurement cadence, finding that $f \lesssim 10\,\mathrm{Hz}$ for essentially any reasonable choice of external $B$-field gradient.

\subsection{Projected Sensitivities}

\begin{figure*}[!t]  
\hspace{0pt}
\vspace{-0.2in}
\begin{center}
\includegraphics[width=0.95\textwidth]{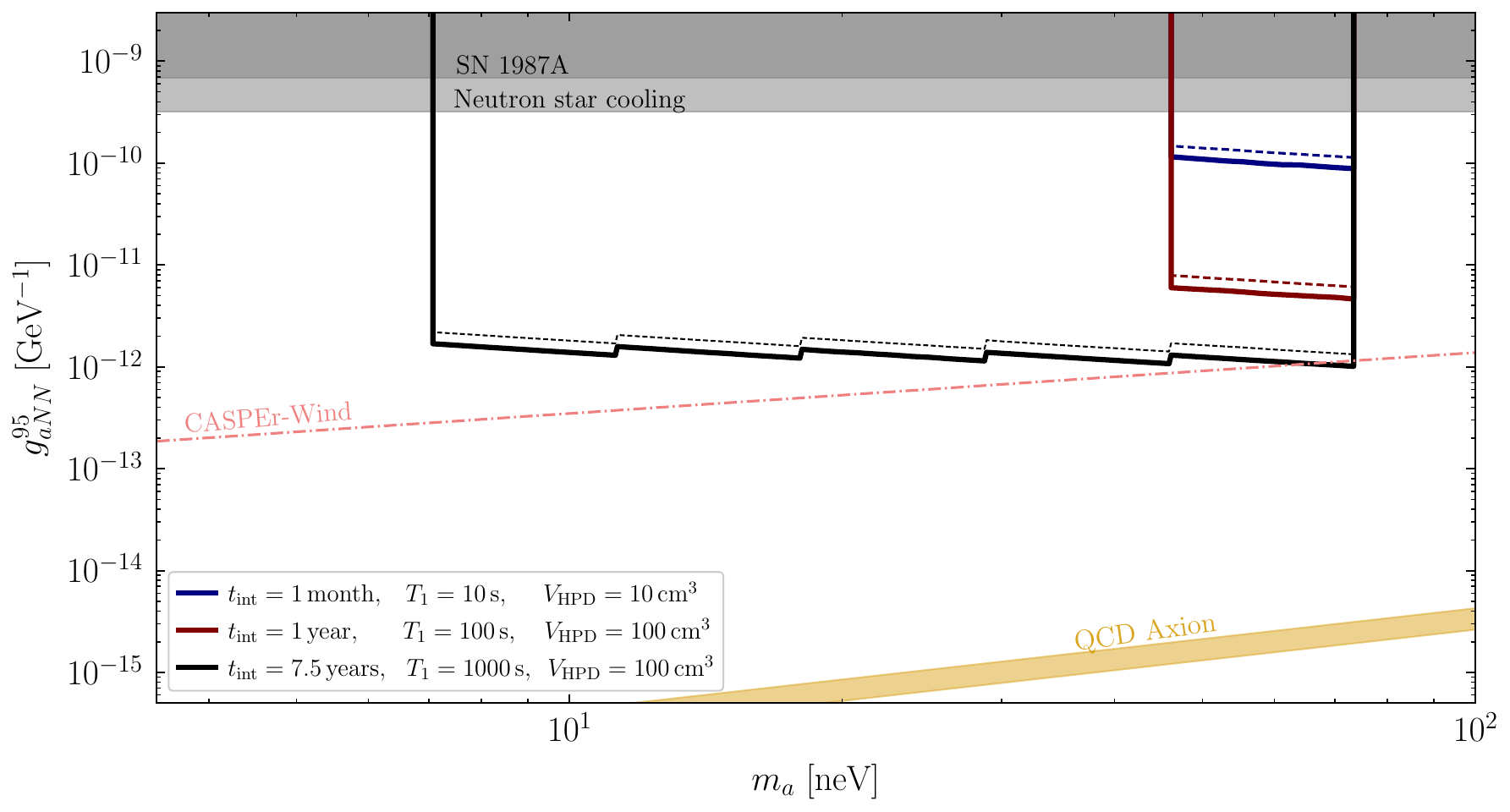}
\caption{Projected 95$^\mathrm{th}$ percentile upper limits on $g_{aNN}$ for three benchmark scenarios (black, red, blue, in order of most to least aggressive) which vary the total collection time, relaxation time, and system volume. In all scenarios, we fix $z_0(t_0) = h = 10\,\mathrm{cm}$ and $\alpha = 1\,\mathrm{cm}^{-1}$. In solid lines, we assume a measurement precision provided by $N_q = 100$ qubits, each with a coherence time of $1\,\mathrm{ms}$, while in dashed lines, we assume only $T_q = 0.1\,\mathrm{ms}$. Otherwise identical parameters are used. We also indicate existing constraints from SN 1987A~\cite{Carenza_2019} and neutron star cooling~\cite{Beznogov_2018} as well as projected constraints for CASPEr-Wind from \cite{JacksonKimball:2017elr}. The gold band indicates the expected couplings for the QCD axion~\cite{Graham:2013gfa}.}
\label{fig:Projection}
\end{center}
\end{figure*}

We collect the results developed in Sec.~\ref{sec:ClockSensitivity} and Sec.\ref{sec:GradSensitivity} to develop projections for extended HPD measurements in search of axion dark matter. For our frequency measurement scheme, we assume the cryogenic sapphire microwave oscillator clock standard (a technology which presently exists in commercial form) rather than a higher-performing optical clock that may be viable in the future. We assume a total height of the HPD system of $h = 10\,\mathrm{cm}$, and we choose $\alpha = 1\,\mathrm{cm}^{-1}$ so that $\alpha h \approx 10$. Note that this implies that the magnetic field varies by an order of magnitude between the top and bottom of the sample container. When then consider three collection scenarios, calculating the sensitivity assuming $N_q = 100$  for both $T_q = 0.1\,\mathrm{ms}$ and, more optimistically, $T_q = 1 \,\mathrm{ms}$.

In the first and most conservative, we assume an HPD system with a total volume of $10\,\mathrm{cm}^3$ with a relaxation time of $T_1 = 10\,\mathrm{s}$. We choose $B_0 = 0.05 \,\mathrm{T}$ so that the field strength over the HPD height varies between $0.05\,\mathrm{T}$ and $0.55\,\mathrm{T}$; as noted in Ref.~\cite{Gao:2022nuq}, stronger field strengths would destabilize the HPD~\cite{scholz1981magnetic}. We assume that for a period of $t_{\rm int} = 1$ month, the HPD system is prepared with the domain wall at a height of $h = 10\,\rm{cm}$, then allowed to relax for time $T_1$ before being re-prepared. The projected sensitivity for $T_q = 1 \ {\rm ms}$ ($T_q = 0.1 \ {\rm ms}$) are shown in solid (dotted) blue in Fig.~\ref{fig:Projection}. The one-month projected sensitivities are obtained by rescaling the single scan coupling sensitivity by $(1\,\mathrm{month}/10 \,\mathrm{s})^{1/4}$. Even this modest setup can lead to world-leading constraints on the axion-nucleon coupling $g_{aNN}$, exceeding supernova~\cite{Carenza_2019} and neutron star cooling~\cite{Beznogov_2018} constraints.

In a more optimistic scenario, we take a HPD volume of $100\,\mathrm{cm}^3$ with a relaxation time of $T_1 = 100\,\mathrm{s}$. Again, we take $B_0 = 0.05\,\mathrm{T}$, but now allow for a total collection time of $t_{\rm int} = 1.0$ years with an otherwise identical data-collection strategy. The projected sensitivity curves are shown in red in Fig.~\ref{fig:Projection}. Finally, to estimate the ultimate sensitivity of an HPD axion experiment, we assume an HPD volume of $100\,\mathrm{cm}^3$ with a relaxation time of $T_1 = 1000\,\mathrm{s}$. We assume $t_{\rm int} = 7.5$ years of total collection time shared equally between 5 different B-field configurations set by $B_0^i = 0.55 \,\mathrm{T} \times (1 + \alpha h)^{-i}$, $i = 1, 2, \dots, 5$, leading to a competitive projected sensitivity across one decade of axion masses. Our choice of 7.5 years is motivated by matching the collection time of CASPEr-Wind over an equal mass range \cite{Graham:2013gfa, Dror:2022xpi}. The projected sensitivities for both measurement cadences are shown in black in Fig.~\ref{fig:Projection}. In particular, the projected sensitivity of the HPD measurement is comparable to that of CASPEr-Wind~\cite{JacksonKimball:2017elr} at the highest frequencies accessible to the HPD setup.\footnote{New aspects of the sensitivity calculation relevant for CASPEr-Wind were pointed out in \cite{Dror:2022xpi} which enhance the projected sensitivities for $m_a \gtrsim 10\,\mathrm{neV}$ relative to the calculations of \cite{Graham:2013gfa}. Related modifications should be expected for the projected sensitivities of \cite{JacksonKimball:2017elr}, but depend in detail on unspecified experimental parameters and scanning strategies.}

\section{Temporal Modulation Effects} 
\label{sec:Modulation}
An important feature of DM experiments is that the Earth's rotation and revolution have the effect of making the lab-frame velocity distribution -- and therefore the signal covariance -- time-dependent. To account for these phenomena, we revisit our expression in Eq.~\ref{eq:FisherInformation}, promoting the signal covariance to be time-dependent as 
\begin{equation}
    \sigma_{A}^{-2} = \sum_n \mathrm{Tr} \left[ (\boldsymbol{S}_n \boldsymbol{B}^{-1})^2 \right]
\end{equation}
where $i$ indexes the time of the collection, $\boldsymbol{S}_n$ is the signal covariance at collection time $t_n$, and we have assumed the background covariance is identical across all collections, \textit{i.e.}, $T_1$ is constant. As a benchmark, we choose $m_a = 73\,\mathrm{neV}$, and we take the magnitude of Earth's peculiar velocity in the galactic frame to be $v_\oplus = 220\, \mathrm{km/s}$ and adopt the Standard Halo Model velocity dispersion $v_0 = 155 \mathrm{km/s}$. For our HPD parameters, we use the same fiducial parameters from Sec.~\ref{sec:Sensitivity}: $\alpha = 1\,\mathrm{cm}^{-1}$, $T = 1000\,\mathrm{s}$, $B_0 = 0.05\, \mathrm{T}$, and $z_0(t_0) = h = 10\,\mathrm{cm}$. We also assume the cryogenic sapphire microwave oscillator clock standard with $N_q = 100$ and $T_q = 0.1\,\mathrm{ms}$ with an optimized sampling cadence.

\begin{figure*}[ht]  
\includegraphics[width=0.99\textwidth]{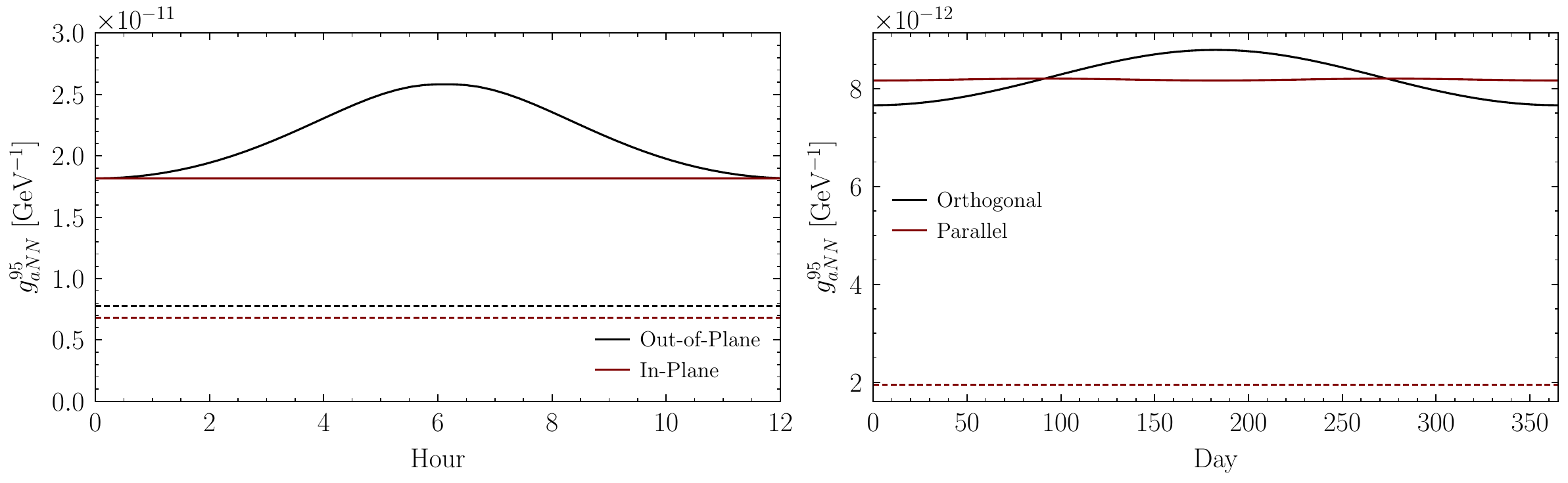}
\caption{(\textit{Left}) A comparison of the effects of daily modulation for two orientations of the rotational axis of Earth. In solid black, we show the modulated sensitivity for 1 hour of collection time expected for an orientation such that the Earth's peculiar velocity rotates from fully-in the domain wall plane to fully orthogonal over a six-hour period. In solid red, we show the analogous modulation expected for the more modest scenario in which the Earth's peculiar velocity rotates only within the plane of the domain wall. Such a modulation effect can only be detected if the phase under the background evolution is well-known at times well-before the axion-induced precession rate shift becomes relevant. Dashed black and dashed red indicate the sensitivity integrated over a 24-hour collection period, where the difference has become marginal. (\textit{Right}) In solid lines, the 1-day sensitivities in our orthogonal and parallel scenarios as a function of the date. In dashed lines, the sensitivity of those collections integrated over 1 year of data-taking.}
\label{fig:Modulation}
\end{figure*}

We first consider two daily modulation scenarios, corresponding to different orientations of the experimental apparatus. In the ``out-of-plane'' scenario, we parametrize the daily modulation as
\begin{equation}
    \mathbf{v}_\oplus = v_\oplus \left[\cos\left(\frac{2\pi t}{24\,\mathrm{hours}}\right)\hat{\mathbf{x}} + \sin\left(\frac{2\pi t}{24\,\mathrm{hours}}\right)\hat{\mathbf{z}} \right],
\end{equation}
where the peculiar velocity rotates in and out of the domain wall plane. In the ``in-plane'' scenario, we instead have 
\begin{equation}
    \mathbf{v}_\oplus = v_\oplus \left[\cos\left(\frac{2\pi t}{24\,\mathrm{hours}}\right)\hat{\mathbf{x}} + \sin\left(\frac{2\pi t}{24\,\mathrm{hours}}\right)\hat{\mathbf{y}} \right],
\end{equation}
where the peculiar velocity rotates only within the plane of the domain wall. The sensitivities to the axion coupling for these two scenarios, both instantaneously and integrated over a full 24-hour data-collection period, are shown in the left panel of Fig.~\ref{fig:Modulation}. As expected, in the out-of-plane scenario, the sensitivity is worst when the peculiar velocity is orthogonal to the domain wall as this minimizes the magnitude of the transverse axion gradient. In the in-plane scenario, there is a small modulation associated with our choice in the expansion of $\theta_0(t)$ to set $\theta = 0$ at the time of when $\dot\theta_0(t) = m_a$, which picks a preferred direction for the system. This modulation would be inaccessible in a realistic statistical analysis if $\theta_0$ could not be determined to sufficient accuracy. In total, the effect of daily modulation appears to be, at most, an $\mathcal{O}(50\%)$ effect on the sensitivity to $g_{aNN}$, and the best sensitivity is achieved in the in-plane scenario when the magnitude of the axion gradient is maximized at all collection times.

We take a similar approach to studying the effect of annual modulation. Holding all HPD and axion parameters fixed, we again consider two scenarios. Taking the speed associated with Earth's revolution velocity around the Sun, $v_\mathrm{rev} = 30\,\mathrm{km/s}$, and fixing the peculiar velocity to lie in the $\hat{\mathbf{x}}$ direction, we consider a ``orthogonal" scenario 
\begin{equation}
    \mathbf{v}_\oplus = v_\oplus \hat{\mathbf{x}} +  v_\mathrm{rev}\left[\cos\left(\frac{2\pi t}{365\,\mathrm{days}}\right)\hat{\mathbf{x}} + \sin\left(\frac{2\pi t}{365\,\mathrm{days}}\right)\hat{\mathbf{z}} \right]
\end{equation}
in which the rotational axis of the Earth's revolution is orthogonal to the peculiar velocity of the Earth. Alternatively, we have a ``parallel" scenario in which 
\begin{equation}
    \mathbf{v}_\oplus = v_\oplus \hat{\mathbf{x}} + v_\mathrm{rev}\left[\cos\left(\frac{2\pi t}{365\,\mathrm{days}}\right)\hat{\mathbf{y}} + \sin\left(\frac{2\pi t}{365\,\mathrm{days}}\right)\hat{\mathbf{z}} \right].
\end{equation}
wherein the rotational axis of the Earth's revolution is parallel to its peculiar velocity. Experimentally, these two scenarios would correspond to rotating the experimental apparatus on a daily basis in order to maintain the relative orientation of the peculiar velocity. The axion-coupling sensitivity of each of these scenarios is shown in the right panel of Fig.~\ref{fig:Modulation}. The bulk of the sensitivity comes from the component of the gradient along the peculiar velocity direction $\hat{\mathbf{x}}$ direction. In the orthogonal scenario, modulation of this boost leads to $\mathcal{O}(10\%)$ changes in the sensitivity over the course of the year. In the parallel scenario, there is no modulation of the boost in the $\hat{\mathbf{x}}$ direction, and so the sensitivity is very nearly constant. Over the span of a year, these modulation effects amount to a sub-percent change in the integrated sensitivity.

It is perhaps surprising that the ``smoking gun'' signature of a DM daily modulation has such a mild effect on the sensitivity to axions. In fact, this behavior is characteristic of searches for wave-like dark matter, where a modulation is an excellent tool for \emph{confirming} a signal, but a mediocre one for \emph{excluding} a signal. Indeed, as noted in Ref.~\cite{Foster:2020fln}, daily modulation is a powerful statistical tool in the sense that once a putative signal is seen, its DM interpretation can often be confirmed at the same level of statistical confidence within a few days. The same holds true for our HPD setup. To achieve the best possible exclusion, though, our analysis shows that one should align the sample in the ``in-plane'' daily modulation scenario to the extent possible.

\section{Conclusion}
\label{sec:Conclusion}

In this paper, we have extended the proposal of Ref.~\cite{Gao:2022nuq} for axion wind detection with the $\He$ HPD with a full statistical treatment of both the signal and background. In doing so, we have quantified the effect of clock noise and measurement error, and developed a data-taking strategy closely related to one which has been proven to be optimal for measuring a frequency of a rotating magnetic field. We have also identified optimal experimental characteristics in order to maximize overall sensitivity, including order-1 external $B$-field gradients and alignment of the sample with the Earth's peculiar velocity. Taken together, these optimizations lead to projected exclusion limits that are competitive with the CASPEr-Wind experiment for the same data-taking time. More generally, we have, to our knowledge, presented the first time-domain formalism for analysis in search of transient axion signals (though in our case, the transient nature comes from the detection medium, rather than the axion source).

In our measurement scheme, the scaling of sensitivity with qubit parameters is perhaps peculiar from the standpoint of quantum information. The measurement noise is relatively insensitive to qubit coherence time, but depends strongly on the SQUID-qubit coupling, so we could likely afford a reduction in coherence time if the benefit were stronger coupling. Likewise, the qubit multiplicity is simply used to reduce spin projection noise, and the qubits are read out independently, so scaling to a large number of qubits on a chip may be easier than quantum computing applications which require the qubits to remain entangled and coherent. Regardless, our experiment is highly synergistic with improvements in superconducting qubit technology, as increases in $T_q$ and $N_q$ can help reduce noise down toward the magnon noise floor.

We emphasize, though, that even with all statistical details and realistic sources of measurement error included, our work confirms the expectation of Ref.~\cite{Gao:2022nuq}: world-leading limits, surpassing all astrophysical bounds, can be achieved with a month of data-taking using commercially-available frequency references and HPD and qubit parameters which can reasonably be achieved in the laboratory. Compared to the axion-photon coupling (see Ref.~\cite{Irastorza:2018dyq} for a review), the axion-nucleon coupling parameter space is quite under-explored, and we hope our proposal provides a path forward for rapid progress in this domain.

\begin{acknowledgments}
{\it 
We thank Jon Ouellet, Nick Rodd, and Yue Zhao for helpful discussions. We thank Mikael Backlund, Brian DeMarco, Elizabeth Goldschmidt, and Wolfgang Pfaff, as well as Dan Carney, Andrei Derevianko, Andrew Jordan, John Howell, Kater Murch, and the other participants of the ``New Directions in Quantum Metrology'' workshop at the Kavli Institute for Theoretical Physics for enlightening conversations on precision measurement strategies. Y. Kahn thanks the Kavli Institute for Theoretical Physics (supported in part by the National Science Foundation under Grant No.\ NSF PHY-1748958) for hospitality during the completion of this work. J. Foster was supported by a Pappalardo fellowship. The work of Y. Kahn and J. Sch\"utte-Engel was supported in part by DOE grant DE-SC0015655. C. Gao was supported by the DOE QuantISED program through the theory consortium “Intersections of QIS and Theoretical Particle Physics” at Fermilab. WH, MN, and JWS acknowledge support from the NSF Division of Materials Research grant DMR-2210112. This  material  is  based  upon  work  supported  by  the U.S.\ Department of Energy,  Office of Science,  National Quantum  Information  Science  Research  Centers,   Superconducting  Quantum  Materials  and  Systems  Center (SQMS) under contract number DE-AC02-07CH11359.
}

\end{acknowledgments}
\appendix

\section{Gaussian Statistics of the Gradient Field}
\label{app:Gaussian}

\begin{figure*}[t!]  
\hspace{0pt}
\vspace{-0.2in}
\begin{center}
\includegraphics[width=0.99\textwidth]{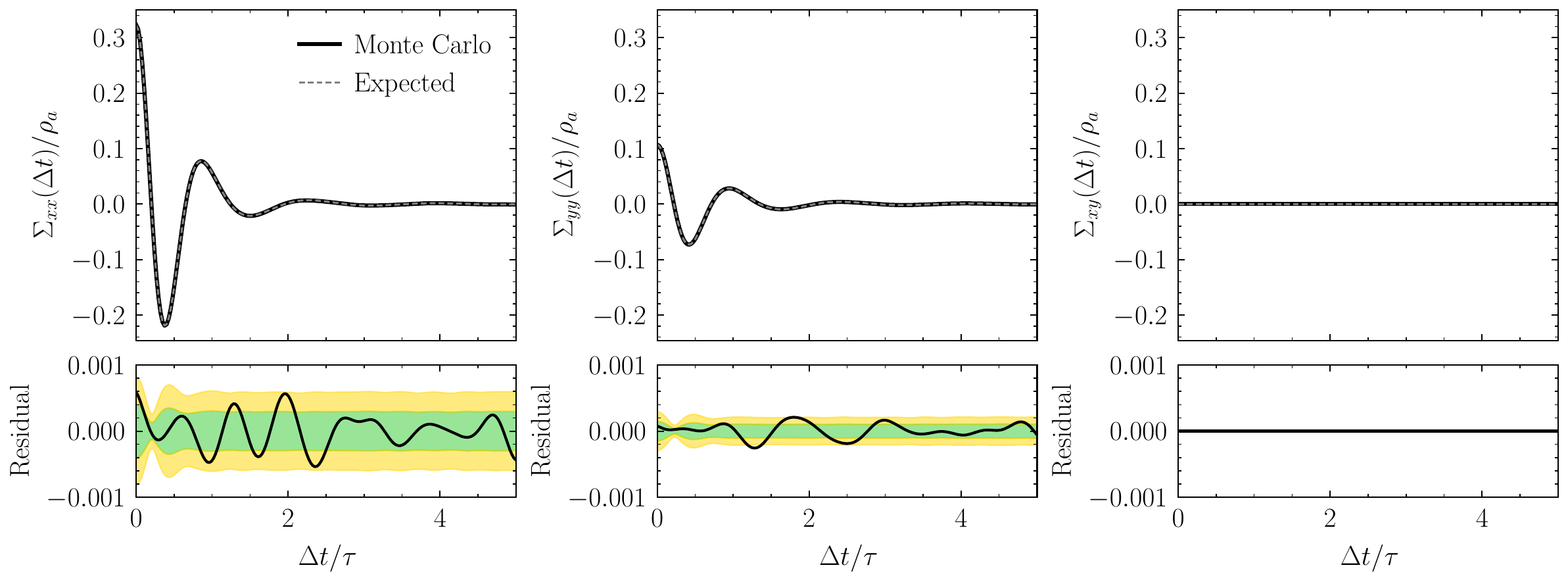}
\caption{A comparison of the correlations $\Sigma_{xx}$ (left), $\Sigma_{yy}$ (center), and $\Sigma_{xy}$ (right) at arbitrary lag $\Delta t \equiv t - t'$ associated with the $\hat{x}$ and $\hat{y}$ components of the axion gradient field as realized by an ensemble of 2,560 Monte Carlo simulations, compared with analytic calculation.  (\textit{Above}) In black, the correlation realized in Monte Carlo; in dashed grey, the analytic expectation. (\textit{Below}) In black, the difference between the Monte Carlo and analytic calculations, with $1\sigma$ (green) and $2\sigma$ (yellow) containment intervals determined from the Monte Carlo ensemble. The axion coherence time is $\tau \equiv 1/(m_a v^2)$, and so these results demonstrate good agreement between analytic expectation and the Monte Carlo for multiple coherence times. The $\Sigma_{xy}$ correlator (right panel) is analytically expected to be identically zero, which is realized at floating point precision in the Monte Carlo.}
\label{fig:GradientStats}
\end{center}
\end{figure*}

In this Appendix, we validate the statistical characterization of the axion gradient field developed in Sec.~\ref{sec:GradientStats} using a Monte Carlo procedure along the lines of that in Ref.~\cite{Foster:2017hbq}. In particular, we construct a time-series realization of an axion gradient field at a single spatial point by performing the sum 
\begin{equation}
    \nabla a(t) = \sum_i^{N_a} \sqrt{\frac{2 \rho_a}{N_a}} \cos\left[m_a\left(1 + \frac{\mathbf{v}_i^2}{2}\right)t + \phi_i\right] \mathbf{v}_i
\end{equation}
where $N_a$ is the number of axion wavemodes included in the sum, $\mathbf{v}_i$ is a velocity drawn from the Standard Halo Model, and $\phi_i$ is a phase uniformly drawn between 0 and $2\pi$. As in Refs.~\cite{Foster:2017hbq, Foster:2020fln}, we unphysically increase the SHM velocity distribution parameters by a factor of 1000 for computational simplicity. We then construct the time-series by summing over $4 \times 10^{5}$ wavemodes, with a total collection time which is 50 times that of the axion coherence time at a resolution of $\Delta t = 1/(10 m_a)$. We use the definition of coherence time given in Eq.~(\ref{eq:CoherenceTime}), taking $|v_\mathrm{obs}| + |\sigma_v|$ as the characteristic speed. An essentially identical procedure was utilized in Ref.~\cite{Gramolin:2021mqv}, but with a focus on instead validating the frequency-domain statistics of the gradient signal, while we focus here on time-domain statistics.\footnote{Ref.~\cite{Lisanti:2021vij} also developed time-domain statistics, but for the dot-product of the gradient field with a precessing dipole, and focusing on the modulation of the signal when the axion coherence time is much larger than 1 year.} 

By creating many realizations of the axion gradient field, we can measure with good precision the correlation functions $\Sigma_{ij}(t,t')$ as defined in Eq.~(\ref{eq:GradCov}). Noting that $\Sigma_{ij}(t,t')$ depends only on $\Delta t \equiv t-t'$, we measure the correlation at arbitrary lag $\Delta t$ from 2,560 Monte Carlo realizations, which are presented in Fig.~\ref{fig:GradientStats} for $\Sigma_{xx}$, $\Sigma_{yy}$ and $\Sigma_{xy}$. Our simulations show excellent agreement with analytic expectations. We obtain similar agreement between the ensemble of Monte Carlo realizations and analytic expectations for correlations involving the $\hat{z}$ component, which we do not present here in the interest of brevity.

\section{Domain Wall Motion as a Stochastic Dynamical System}
\label{App:SDE}
In the main text, we have argued that stochasticity in the magnon loss does not significant affect the resonant dynamics of the HPD-axion system. In this appendix, we formulate the dynamics of the HPD as a system of stochastic differential equations (SDEs) so that we may validate that claim.

We begin by considering the dynamics of the system in the absence of an axion gradient field. The It\^{o} SDE describing the exponential decay of magnons in the system is given by
\begin{equation}
    dN_d(t) = -\Gamma N(t) dt + \left[\Gamma N(t)\right]^{1/2} dW_t
    \label{eq:StochasticDecay}
\end{equation}
where $\Gamma$ is the instantaneous decay rate and $dW_t$ is a standard Wiener process. 

With insight from the Bloch equations describing the effective dynamics of the HPD, we note that the transverse components of the magnetization may be directly evaluated from the state variables $N$ and $\theta$, and we denote those transverse magnetizations by $M_x(N, \theta)$ and $M_y(N, \theta)$. The instantaneous rate of change of $M_z$ in the presence of an axion gradient field is then
\begin{equation}
    dM_z = \gamma \left[B_x^a(t) M_y(N, \theta) - B_y^a(t) M_x(N, \theta)\right] dt.
\end{equation}
The magnetization can be related to the total magnon number, so that the instantaneous rate of change in the total magnon number due to the axion is given by 
\begin{equation}
dN_a(t) = -\frac{5 h A}{\gamma} \left[B_x^a(t) M_y(N, \theta) - B_y^a(t) M_x(N, \theta)\right] dt
\label{eq:AxionSource}
\end{equation}
where $h$ is the height of the sample container and $A$ is its cross-sectional area. We combine these results to obtain
\begin{equation}
    dN(t) = dN_d(t) + dN_a(t)
    \label{eq:dN}
\end{equation}
describing the evolution of the magnon number in the HPD system. 

As in Sec.~\ref{sec:StochasticDynamics}, for a given $N(t)$ and known magnetic field profile $B(z)$, we may evaluate the domain wall height $z$ as a function of $N$, which we denote $z(N)$. Since the precession frequency $\omega_L$ is determined solely by the magnetic field strength at the domain wall, we have the ordinary differential equation (ODE) for the phase
\begin{equation}
    d\theta(t) = \gamma B[z(N(t))] dt.
    \label{eq:dTheta}
\end{equation}
Together the system of the SDE in Eq.~(\ref{eq:dN}) and the ODE in Eq.~(\ref{eq:dTheta}) may be be straightforwardly solved using standard libraries, \textit{e.g.}, \texttt{diffrax} to study realizations of the stochastic dynamics of the HPD system \cite{kidger2022neural}.

\subsection{Toy Parameters for Computationally Tractable Simulations}

There remain computational challenges associated with numerical simulation of the stochastic dynamics. For realistic experimental parameters, the magnon number in the HPD is very large, $\mathcal{O}(10^{20})$. Even with the unphysical DM velocity $v \sim 0.1$, accurately evolving the equations of motion requires that we use a timestep $\Delta t \lesssim 1 / (10 m_a) \approx 10^{-7}\, \mathrm{s}$. For $T_1 = 1000\, \mathrm{s}$, the magnitude of the diffusion term in the SDE is $\mathcal{O}(10^{10})$. Hence, the typical change in the magnon number associated with stochastic decay is more than 16 orders of magnitude below the magnon number, which is below floating-point precision. As a result, our simulations will use physically-unrealistic parameters that reduce this hierarchy to be within floating-point precision.\footnote{Alternatively, a implementation of an SDE-solver at long double precision would suffice.}

As in App.~\ref{app:Gaussian}, we perform an unphysical rescaling of the HPD parameters in order to bring the system into a computationally tractable regime. In these toy simulations, we reduce $\chi$ from its realistic value of $10^{-7}$ to $10^{-15}$ and reduce the gyromagnetic ratio by a factor of $1000$ by $\gamma \rightarrow \tilde \gamma = \gamma/1000$.
In total, this reduces the hierarchy between the scatter in stochastic decay and the magnon number to roughly 8 orders of magnitude, which is well within floating-point precision. Reducing the precession frequency through the rescaling of $\gamma$ has the additional advantage of increasing the step size $\Delta t$, making computations over a fixed time interval considerably faster.

For our subsequent comparisons of the fully nonlinear SDE developed in this section and the linearized, fully deterministic description developed in the main text, we take $z_0(t_0) = h = 10\,\mathrm{cm}$, $T_1 = 1000\,\mathrm{s}$, $\alpha =1.0\,\mathrm{cm}^{-1}$, and $B_0 = 0.05\,\mathrm{T}$. We choose an axion mass $m_a \approx 73\,\mathrm{peV}$ so that the resonance $m_a = \dot \theta$ is expected at $t_r \approx 5\,s$. For simplicity, we take the axion gradient field to be monochromatic and point only in the $\hat{\mathbf{x}}$ direction:
\begin{equation}
    B_x^a(t) = \frac{g_{aNN} v \sqrt{2 \rho_a}}{\gamma} \cos(m_a t) , \qquad B_y^a(t) = 0,
\end{equation}
with $v = 300\,\mathrm{km/s}$. For our toy model parameters, a relatively large value of $g_{aNN} = 1.0 \times 10^{-9} \,\mathrm{GeV}^{-1}$ is necessary in order to produce an observable effect. Though this is not physically realistic (it is already excluded by astrophysical constraints, see Fig.~\ref{fig:Projection}), it suffices to demonstrate the effectiveness of the linear deterministic framework.

\subsection{Realizations of Nonlinear SDE and Linear ODE Descriptions}
\label{App:MonochromaticExample}

\begin{figure}[t!]  
\includegraphics[width=0.49\textwidth]{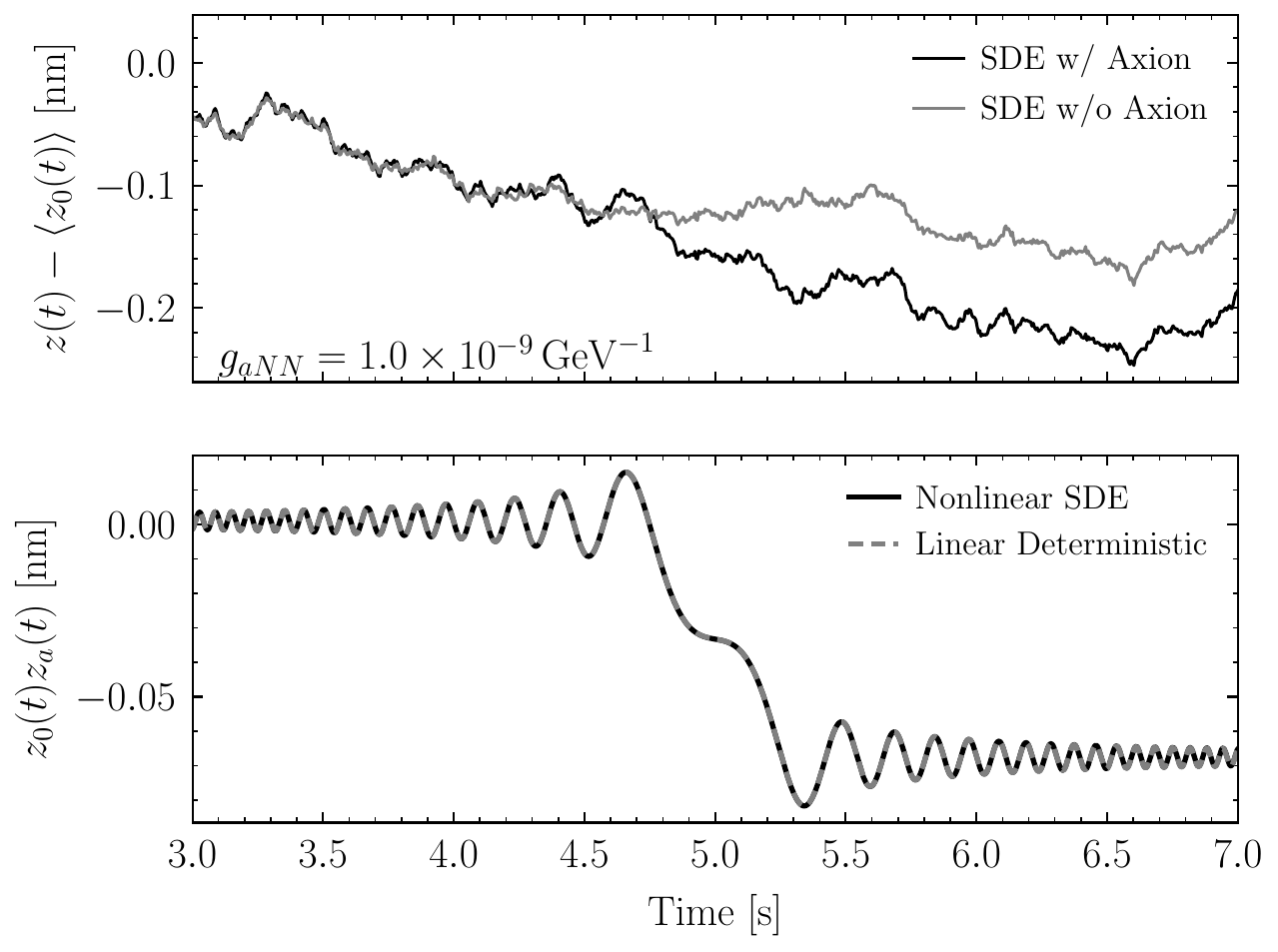}
\caption{(\textit{Top}) A comparison of the difference between the stochastic evolution of the domain wall position with and without the axion gradient and the expected domain wall position in absence of an axion gradient. (\textit{Bottom}) A comparison of the axion-induced of the axion-induced shift in the domain wall position computed within the nonlinear SDE framework of this Appendix and the linear deterministic ODE framework developed in Sec.~\ref{sec:Dynamics}. For details, see text.}
\label{fig:SDE_Test}
\end{figure}

\begin{figure*}[!t]  
\hspace{0pt}
\vspace{-0.2in}
\begin{center}
\includegraphics[width=0.9\textwidth]{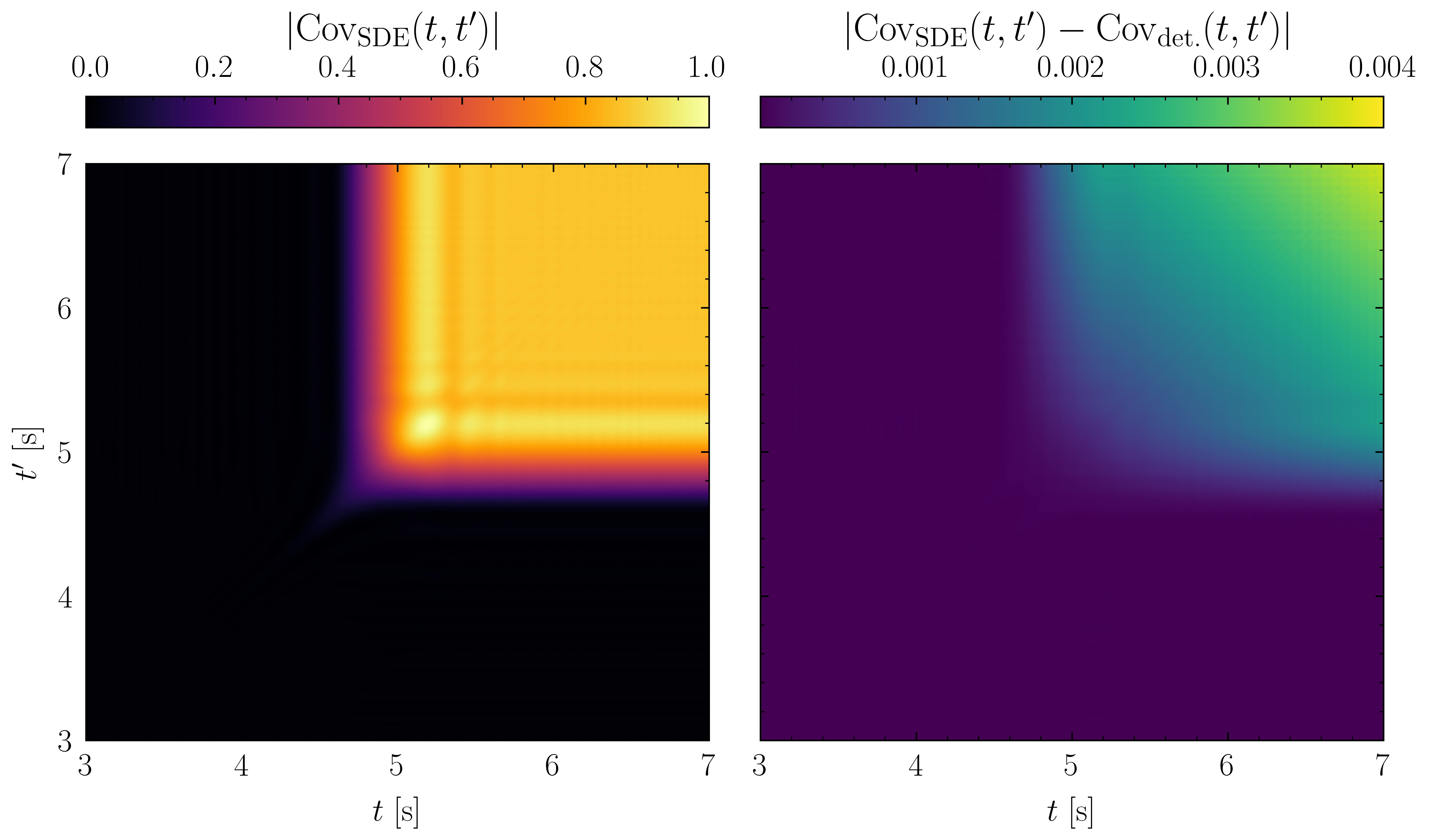}
\caption{(\textit{Left}) The covariance matrix calculated from 512 realizations of axion-induced domain wall motion in the SDE treatment. The covariance matrix is normalized to have maximum value $1$. (\textit{Right}) The difference between the covariance matrices calculated in the SDE and linear deterministic treatment, subject to an identical normalization. Good agreement is observed over the range of relevant times.}
\label{fig:CovTest}
\end{center}
\end{figure*}

In Fig.~\ref{fig:SDE_Test}, we show the evolution of the domain wall position, comparing the linear deterministic ODE framework developed in the main text and the stochastic nonlinear ODE framework developed in this Appendix. In the top panel of Fig.~\ref{fig:SDE_Test}, we show the difference between the a realization of the stochastic domain wall motion and the expected (deterministic) domain wall position in two scenarios. In the first (grey), we assume no axion gradient field so that the difference between the realization and the expected solution is purely due to stochasticity in magnon loss. In the second (black), we include the effect of a realization of an axion gradient field so that the difference between the realization and the expectation is due to the cumulative effect of the gradient field and stochastic magnon loss. Both realizations are solved using an identical seed for the pseudorandom number generator used in the diffusion (stochastic magnon loss) term. The order-1 difference between the black and grey curves in the neighborhood of the resonance clearly shows the effect of the axion source.

The bottom panel of Fig.~\ref{fig:SDE_Test} plots $z_0(t) z_a(t)$, which is the axion-induced domain wall motion. For the SDE treatment (solid black), we evaluate this quantity from the difference between the realizations which have identical magnon loss but are evaluated with and without an axion gradient field, as shown in the top panel. This difference is then the contribution to the domain wall motion induced by the axion gradient field including all effects associated with the interplay between axion dynamics and stochastic magnon loss. We can also evaluate this quantity in our linear deterministic treatment by taking $z_0(t) = \langle z_0(t)\rangle$; this result is presented in dashed grey. 

The excellent agreement between the two approaches confirms that our linear deterministic approach accurately captures the dynamics of the axion-induced domain wall motion and hence the precession frequency shift. This agreement is remarkable despite the fact that the axion-induced shift in our simulations is even larger than the stochastic shift, \textit{i.e.} we are simulating the regime in which a signal would be highly detectable. For physical parameter values, such as larger values of $\chi$, the magnon number will be larger, further reducing the size of the diffusion term relative to the expected drift. Moreover, for larger values of $m_a$, the duration over which the axion gradient can drive non-negligible domain wall motion is shorter. Hence the size of the relevant stochasticity will be smaller. 

Taken as a whole, we interpret our numerical results as strong evidence for the accuracy of our approximations in Secs.~\ref{sec:Dynamics}--\ref{sec:StochasticDynamics}. Indeed, in the evaluation of the SDE, no approximations regarding slowly varying or rapidly oscillating terms was made. In particular, in our derivation of the HPD equations of motion in Sec.~\ref{sec:Dynamics}, we neglected bare $\cos \theta$ and $\sin \theta$ terms as they oscillate rapidly and average to zero on the finite interval over which measurements are made; our simulation results validate that approximation for our unrealistic parameters, and for realistic parameters, the precession frequency is 1000 times larger, making this an even better approximation. Similarly, in Sec.~\ref{sec:StochasticDynamics}, we argued that terms that depend on $\delta z_0$ in Eq.~(\ref{eq:FullCov}) would be suppressed relative to those that depend on $\langle z_0 \rangle$ by $\delta t / T_1$. Once again, the good agreement in our simulations validate this approximation over a time interval 100 times larger than for physical parameters; thus, for physical parameters, neglecting terms at order $\delta z_0 / \langle z_0 \rangle \approx \delta t /T_1$ is an even better approximation than we are able to depict in this unphysical example.

\subsection{Signal Covariance from SDEs}

We now extend our toy model beyond that of a monochromatic velocity distribution, though for continued simplicity, we will take the distribution to only have support in the $\hat{\mathbf{x}}$ direction. In particular, we take
\begin{equation}
    f(v_x) = \frac{1}{\sqrt{2 \pi \sigma_v^2}}\exp \left[- \frac{(v_x-v_\mathrm{obs})^2}{2 \sigma_v^2} \right]
\end{equation}
with $\sigma_v = 1550\,\mathrm{km/s}$ and $v_\mathrm{obs} = 2200 \,\mathrm{km/s}$, enabling us to generate the time-resolved gradient field directly from its defining covariance and use this as an input for our integration of the domain wall motion. 

We then repeat our procedure from App.~\ref{App:MonochromaticExample} over many independent realizations of the axion field and the stochastic magnon loss. From 512 realizations, we construct a time-time covariance matrix, which we compare to one calculated via our linear deterministic treatment. In Fig.~\ref{fig:CovTest}, we present a summary of these realizations, finding the difference in the covariance matrix between the SDE treatment and the linear deterministic treatment to agree at the sub-percent level.

\section{Measuring Drifting Frequencies with Optimal Control}
\label{sec:OptimalControlMeasurement}

Here, we demonstrate how an optimal control scheme may be utilized to make a measurement of a drifting frequency. Consider a time-dependent Hamiltonian for a two-level system:
\begin{equation}
H_{\dot{\omega}}(t)=A\sin(\omega t+\dot\omega t^2/2)\frac{\sigma_z}2,
\label{eq:Homegadot}
\end{equation}
where the subscript emphasizes that we consider $H$ as a function of the parameter $\dot{\omega}$. We are interested in the fundamental limit on the uncertainty $\delta\dot\omega$ of our desired parameter $\dot{\omega}$, given some measurement time $T$ and a perfect knowledge of $\omega$. This is related to the quantum Fisher information (QFI) which can be formulated as
\begin{equation}\label{eq:I_omega}
I_{\dot\omega}^{(Q)}= \left( \int_0^T(\mu_+(t)-\mu_-(t))dt\right)^2,
\end{equation}
where $\mu_\pm$ are the maximum and minimum eigenvalues of the operator $\partial_{\dot\omega} H_{\dot\omega}$ : 
\begin{equation}
\label{eq:mu}
\mu_{\pm}(t)=\pm \frac{At^2}4\cos(\omega t+\dot\omega t^2/2).
\end{equation}
It can be show that without any Hamiltonian control, the QFI scales as $T^2$ for large $T$.

Following the treatment of Ref.~\cite{naghiloo2017achieving}, a superposition of eigenvalues of $H_{\dot{\omega}}$ maximizes the QFI:
\begin{equation}\label{eq:state}
|\Psi\rangle_\phi =\frac1{\sqrt2}\left(|0\rangle +e^{i\phi}|1\rangle \right).
\end{equation}
Under the action of $H_{\dot{\omega}}$, the two eigenstates will acquire a relative (time-dependent) phase $\phi_{\omega,\dot\omega}(t)$. The optimal control Hamiltonian $H_c$ will consist of $\pi$ pulses at the antinodes of \eqref{eq:Homegadot}, such that the roles of $\mu_+$ and $\mu_-$ are reversed at each antinode and the QFI integrand is positive-definite:
\begin{equation}
\mu^{H_c}_\pm(t)=\pm \frac{At^2}4 \bigg |\cos(\omega t+\dot\omega t^2/2)\bigg|.
\end{equation}
We note that since the operator structure of \eqref{eq:Homegadot} is independent of $\omega$ and $\dot{\omega}$, with both parameters appearing only in the amplitude of $\sigma_z$, the same state preparation can be used to make the optimal measurements of $\omega$ which feed into the measurement of $\dot{\omega}$. In either case, the QFI under optimal control is given by
\begin{equation}
I_{\dot \omega}^{(Q)} = \left[ \int_0^{T} dt\, \frac{At^2}{4} \left |\cos \left( \omega t + \frac{\dot{\omega} t^2}{2}\right) \right | \right]^2.
\end{equation}
It is convenient to perform a change of variable to $t' = t + b t^2$ with $b \equiv \dot \omega / 2 \omega$. So long as $b t' \ll 1$, the QFI is approximately
\begin{equation}
I_{\dot \omega}^{(Q)} \approx \left[ \int_0^{T} dt'  \frac{At'^2}{4} \left |\cos (\omega t') \right | \right]^2,
\end{equation}
which is a good approximation for typical HPD experiments in which the measurement interval over which $\dot \omega$ is examined is shorter than the relaxation time $T_1$. 

For convenience, we define $T_\omega \equiv 2 \pi /\omega$ and assume the measurement interval $T$ to be an integer multiple $N$ of $T_\omega$. Then the QFI may be rewritten as 
\begin{equation}
\begin{split}
I_{\dot \omega}^{(Q)} \approx \bigg[\sum_{n=1}^N \frac{A}{4}\Big( \int_{(n-1)T_\omega}^{(n-1)T_\omega+\frac14T_\omega} t'^2\cos(\omega t')dt'\\
+\int_{(n-1)T_\omega+\frac34T_\omega}^{n T_\omega}  t'^2\cos(\omega t')dt'\\
-\int_{(n-1)T_\omega+\frac14T_\omega}^{(n-1)T_\omega+\frac34T_\omega}  t'^2\cos(\omega t')dt'\Big)\bigg]^2
\end{split}
\end{equation}
without any absolute values. These integrals are trivial to evaluate and ultimately we find 
\begin{equation}
\label{eq:IFI}
    I_{\dot \omega}^{(Q)} \approx \left(\frac{A T^3}{3 \pi} \right)^2
\end{equation}
at leading order, which is a marked improvement over the $T^2$ scaling in the absence of control.

The QFI is a measure of the distinguishability of two states $|\Psi\rangle_\phi$ and $|\Psi\rangle_{\phi+d\phi}$,  and thus can be given in terms of the Bures distance
\begin{equation}
I_{\dot\omega}^{(Q)}= \frac{4 ds^2}{d\dot\omega^2},
\end{equation}
where $ds^2=2(1-|\langle \Psi_\phi|\Psi_{\phi+d\phi}\rangle|)$. 
Given the state preparation in \eqref{eq:state}, we have
\begin{equation}
ds^2 =2\left(1-\left|\frac12+\frac12 e^{i \delta \phi}\right|\right)\approx \frac{\delta\phi^2}4 +\mathcal{O}(\delta\phi^4)
\end{equation}
and we get for the QFI
\begin{equation}\label{eq:I_2}
I_{\dot\omega}^{(Q)}\approx \left(\frac{\delta\phi}{\delta \dot\omega}\right)^2~.
\end{equation}
Equating \eqref{eq:I_2} with \eqref{eq:IFI}, we have
\begin{equation}
\delta\dot\omega \approx \frac{3\pi}{AT^3}\delta\phi \to \frac{3\pi}{AT^3}\frac1{2\sqrt{N_q}}
\end{equation}
where we used $\delta\phi=1/\sqrt{4N_q}$ as an estimate of the phase noise for $N_q$ independent measurements. In our scheme, the independent measurements may be either the number of repeated experimental measurements or the number of independent qubits. 

This derivation above only holds for a measurement time during which the qubit remains coherent. Following Ref.~\cite{schmitt2017submillihertz}, suppose now the qubit is characterized by a coherence time $T_q$, and measurement of the frequency drift is made over a duration $T=N_m T_q$. Then the QFI accumulates as 
\begin{equation}
I_{\dot\omega}^{(Q)}=\sum_{n=1}^{N_m} \left( \int_{(n-1)T_q}^{nT_q}(\mu_+(t)-\mu_-(t))dt\right)^2.
\end{equation}
To evaluate the QFI, we make the simplifying assumption that $T_q = N T_\omega$ with $N \gg 1$. We then have
\begin{widetext}
\begin{equation}
I_{\dot\omega}^{(Q)} = \left[\frac A2 \int_{(n-1)T_q}^{nT_q}t^2 |\cos(\omega t)| dt\right]^2 =\left[ \frac A2\sum_{n'=1}^N\left(\int_{(n-1)T_q+ (n'-1)T_\omega}^{(n-1)T_q+ (n'-1)T_\omega+\frac14T_\omega}  t^2 \cos(\omega t)dt+\cdots \right)\right]^2 \approx \frac{A^2}{5 \pi^2} T_q T^5
\end{equation}
\end{widetext}
at leading order. Following identical reasoning using the Bures metric to relate the $\delta \phi$ and $\delta \dot{\omega}$, we now find
\begin{equation}
\delta\dot\omega \approx \frac{\sqrt5 \pi}{2A T^{5/2}\sqrt{T_q N_q}}.
\end{equation}
Hence we see the effect of qubit incoherence is to only slightly soften the scaling of our precision with time from $T^{-3}$ to $T^{-5/2}$. \bibliography{refs}
\end{document}